\def\comp{{\rm C}\llap{\vrule height7.1pt width1pt depth-.4pt\phantom t}}
\def\Box{\kern1pt\vbox{\hrule height 1.2pt\hbox{\vrule width 1.2pt\hskip 3pt
   \vbox{\vskip 6pt}\hskip 3pt\vrule width 0.6pt}\hrule height 0.6pt}\kern1pt}
\def\gtwid{\mathrel{\raise.3ex\hbox{$>$\kern-.75em\lower1ex\hbox{$\sim$}}}}
\def\ltwid{\mathrel{\raise.3ex\hbox{$<$\kern-.75em\lower1ex\hbox{$\sim$}}}}
\def\sla{\raise.15ex\hbox{$/$}\kern-.57em}

\documentstyle[epsfig,12pt]{article}

\newcommand{\be}{\begin{equation}}
\newcommand{\ee}{\end{equation}}

\begin{document}
\begin{titlepage}
\begin{flushright}
UFIFT-HEP-02-18 \\ hep-ph/0207190
\end{flushright}
\vspace{.4cm}
\begin{center}
\textbf{The Axial Anomaly in D=3+1 Light-Cone QED}
\end{center}
\begin{center}
M. E. Soussa$^{\dagger}$ and R. P. Woodard$^{\ddagger}$
\end{center}
\begin{center}
\textit{Department of Physics \\ University of Florida \\
Gainesville, FL 32611 USA}
\end{center}

\begin{center}
ABSTRACT
\end{center}
We consider $(3+1)$-dimensional, Dirac electrons of arbitrary mass, 
propagating in the presence of electric and magnetic fields which
are both parallel to the $x^3$ axis. The magnetic field is constant
in space and time whereas the electric field depends arbitrarily 
upon the light-cone time parameter $x^+ = (x^0 + x^3)/\sqrt{2}$.
We present an explicit solution to the Heisenberg equations for 
the electron field operator in this background. The electric field
results in the creation of electron-positron pairs. We compute the 
expectation values of the vector and axial vector currents in the 
presence of a state which is free vacuum at $x^+ = 0$. Both current 
conservation and the standard result for the axial vector anomaly 
are verified for the first time ever in $(3+1)$-dimensional light-cone 
QED. An interesting feature of our operator solution is the fact 
that it depends in an essential way upon operators from the 
characteristic at $x^- = -L$, in addition to the usual dependence 
upon operators at $x^+ = 0$. This dependence survives even in the
limit of infinite $L$. Ignoring the $x^-$ operators leads to a 
progressive loss of unitarity, to the violation of current 
conservation, to the loss of renormalizability, and to an incorrect 
result for the axial vector anomaly.

\begin{flushleft}
PACS numbers: 11.15.Kc, 12.20.Ds, 11.40.Dw
\end{flushleft}
\vspace{.4cm}
\begin{flushleft}
$^{\dagger}$ e-mail: soussam@phys.ufl.edu \\
$^{\ddagger}$ e-mail: woodard@phys.ufl.edu
\end{flushleft}
\end{titlepage}

\section{Introduction}

It is customary in formulating a $(3+1)$-dimensional quantum field theory 
on the light-cone, to regard $x^+ \equiv ( x^0 + x^3)/\sqrt{2}$ as the 
time coordinate. The complimentary null direction, $x^- \equiv \left( x^0 
- x^3 \right)/\sqrt{2}$ is treated as a spatial coordinate, as are the 
transverse variables, $x^{\perp} = (x^1,x^2)$. In this view one is led to 
imagine that the Heisenberg field equations can be solved to express the 
operators at an arbitrary point $(x^+,x^-,x^{\perp})$ in terms of the initial 
value operators on a surface of constant $x^+$.

However, it has been known for some time that solving the Klein-Gordon 
or Dirac equation on the light-cone actually involves initial data on both 
characteristics \cite{HW}. In order to completely determine the operators 
in the wedge with $x^+ > 0$ and $x^- > -L$ one must specify not only
their values for $x^+ = 0$ with $x^- \ge -L$, but also for $x^- = -L$ with 
$x^+ \ge 0$. This remains true even if $L$ is taken to $\infty$ \cite{M1},
although then the problem is segregated to the singularity at $p^+ = 0$. For 
free theories in trivial backgrounds, one can simply constrain this sector 
of the theory. Such a constraint is consistent because there is no mode
mixing for these theories.

Interactions introduce mode-mixing, and it is no longer obvious that 
the $p^+ = 0$ modes can be suppressed consistently. Nontrivial background
fields can also result in mode-mixing and recent results in this context
seem to show conclusively that the $p^+ =0$ modes cannot be ignored. We 
now have explicit and completely general solutions to the Heisenberg 
equations for Dirac electrons in the presence of an electric background 
field which points in the $x^3$ direction and is an arbitrary function of 
$x^+$ \cite{TW1,TW2}. The homogeneous electric field results in $e^+ e^-$ 
pair production in an amazingly simple fashion. Each Fourier mode of fixed 
$k^+$ experiences pair production at the instant when its minimally coupled 
momentum, $p^+(x^+) \equiv k^+ - e A_-(x^+)$, vanishes. At this instant 
the electron field operator suffers a drop in the amplitude proportional 
to the initial value data from the $x^+ = 0$ surface, with the missing 
amplitude being supplied by operators from the surface of constant $x^-$.
Suppressing these other operators leads to a progressive loss of unitarity 
and to violation of current conservation. One also fails to produce the 
standard result for the axial vector anomaly in $1+1$ dimensions \cite{TW2}.

Although the first paper \cite{TW1} applies to an arbitrary dimension, the
operator solution was only valid in the limit $L \longrightarrow \infty$.
Since the limit could only be taken in the distributional sense, the solution
was not sufficient to compute the expectation value of certain fermion
bilinears. It is better to obtain a solution for arbitrary $L$, compute the
expectation value of whatever operator is desired first, and {\it then} take
the large $L$ limit of the resulting $\comp$-number. This was done in the 
second paper \cite{TW2}, but all the calculations were restricted to $1+1$ 
dimensions. In this paper we compute in $3+1$ dimensions. We have also 
extended the background to include a constant magnetic field which is 
co-linear with the electric field. This allows us to check the axial vector 
anomaly for the first time ever in $3+1$-dimensional light-cone QED.

This Introduction is the first of seven sections. Section 2 explains 
light-cone notation and gauge choices. It also presents our solution of 
the Dirac equation in the previously described background. Section 3 
describes quantization and also explains how to work in the presence of 
a state which is empty on the initial value surface. In Section 4 we 
calculate the probability of pair creation. Section 5 is devoted to 
computing the expectation values of the vector currents. In Section 6, we 
show that the expectation values of the axial vector currents $J^{+}_5$, 
$J^{-}_5$, and the pseudoscalar $J_5$ obey the Adler-Bell-Jackiw anomaly 
to all orders in the magnetic field. Section 7 gives concluding remarks.

\section{The model and its solution}

The Lagrangian density for QED is,
\be
{\cal L} = \overline{\Psi}\gamma^{\mu}(i \partial_{\mu} - eA_{\mu} -m)\Psi 
- \frac{1}{4}F_{\mu\nu}F^{\mu\nu}.
\ee
In four dimensions $\mu$ and $\nu$ run from 0 to 3.  $A_{\mu}$ is the gauge 
potential, $\Psi$ is the Dirac bispinor, and $F_{\mu\nu} \equiv \partial_{
\mu}A_{\nu} - \partial_{\nu}A_{\mu}$ is the Maxwell field strength tensor.
We employ the conventions of Bjorken and Drell \cite{BD}, who give
$\eta^{\mu\nu}$ timelike signature and $\{\gamma^{\mu},\gamma^{\nu}\} 
= 2\eta^{\mu\nu}$. 

The coordinates of light-cone quantum field theory are \cite{KS},
\be
{x^{\pm} \equiv \frac{1}{\sqrt{2}}(x^{0} \pm x^{3}) \qquad , \qquad 
{x^{\perp} \equiv (x^{1},x^{2})}} \; . \label{eq:zero}
\ee
Any vector can be expressed in this basis.  For example, the inner product 
of two Lorentz vectors is,
\be
{a^{\mu}b_{\mu} = a^{+}b^{-} + a^{-}b^{+} - a^{\perp} \cdot b^{\perp}} 
\; . \label{eq:one}
\ee
From (\ref{eq:one}) we are able to extract the nonvanishing components of 
the light-cone metric as $\eta^{+-} = \eta^{-+} = -\eta^{11} = -\eta^{22} 
= 1$. Therefore, raising and lowering are accomplished thusly: $a_+ = a^-$, 
$a_- = a^+, a_1 = -a^1, a_2 = -a^2$. Further, the divergence of any 4-vector 
is, ${\partial_{\mu} V^{\mu} = \partial_+ V^+ + \partial_- V^- + 
\nabla_{\perp} \cdot V^\perp}$.

Light-cone gamma matrices satisfy,
\be
{(\gamma^{\pm})^2 = 0 \qquad , \qquad {\{\gamma^{+},\gamma^{-}\} = 2}} 
\qquad , \qquad \{\gamma^i,\gamma^j\} = -2 \delta^{ij} \; .
\ee
Dirac spinors on the light-cone are decomposed by the projectors,
\be
{P_{\pm} \equiv \frac{1}{2}\gamma^{\mp}\gamma^{\pm} = \frac{1}{2}(I \pm 
\gamma^{0}\gamma^{3})}.
\ee
Acting these on the full bi-spinor gives its $+$ and $-$ components,
\be
{\psi_{\pm} = P_{\pm}\Psi \qquad , \qquad {\Psi = \psi_{+} + \psi_{-}}} \; .
\ee

Our electric and magnetic backgrounds are $\vec{E}(x^+,x^-,x^{\perp}) = 
E(x^+) \widehat{x}_3$ and $\vec{B}(x^+,x^-,x^{\perp}) = B \widehat{x}_3$, 
respectively. We fix the gauge with,
\be
A_+(x^+,x^-,x^{\perp}) = 0 \; . 
\ee
We can also impose the surface conditions, 
\be
A_-(0,x^-,x^{\perp}) = 0 \qquad , \qquad A_1(0,0,x^{\perp}) = - 
A_2(0,0,x^{\perp}) \; . 
\ee
In this gauge the nonzero components of the vector potential are,
\be
A_-(x^+) = - \int_0^{x^+} dy E(y) \qquad , \qquad A_{\perp}(x^{\perp}) = 
\frac{B}{2}(x^2 \widehat{x}_1 - x^1 \widehat{x}_2) \; . \label{gaugepots}
\ee

With these conventions the Dirac equation is,
\be
[i\gamma^{+}\partial_{+} + i\gamma^{-}(\partial_{-} + ieA_{-}) + i
\gamma^{\perp} \cdot {\cal D}_{\perp} - m]\Psi(x) = 0,
\ee
where ${\cal D}_{\perp} \equiv \nabla_{\perp} + i e A_{\perp}$ is the 
transverse covariant derivative of QED. Alternately multiplying this 
equation by $\frac12 \gamma^-$ and $\frac12 \gamma^+$ gives two coupled 
equations involving the light-cone spinors:
\begin{eqnarray}
i\partial_{+}\psi_{+}(x) & = & \frac{1}{2}(m + i \gamma^{\perp} \cdot {\cal 
D}_{\perp}) \gamma^- \psi_- , \label{em1} \\
(i\partial_{-} - eA_{-})\psi_{-}(x) & = & \frac{1}{2}(m + i \gamma^{\perp}
\cdot {\cal D}_{\perp}) \gamma^+\psi+ . \label{em2}
\end{eqnarray}
One solves this system by integrating (\ref{em1}) with respect to $x^+$ and
(\ref{em2}) with respect $x^{-}$,
\begin{eqnarray}
\lefteqn{\psi_{+}(x^{+},x^{-},x^{\perp}) = \psi_{+}(0,x^{-},x^{\perp}) - }
\nonumber \\
& & \hspace{1.25cm} \frac{i}{2}(m + i \gamma^{\perp} \cdot {\cal D}_{\perp})
\int_0^{x^+} du \gamma^- \psi_-(u,x^-,x^{\perp}) , \label{eq:it1} \\
\lefteqn{\psi_{-}(x^{+},x^{-},x^{\perp}) = e^{-ieA_{-}(x^{+})(x^{-}+L)}
\psi_{-}(x^{+},-L,x^{\perp}) - } \nonumber \\
& & \hspace{1.25cm}\frac{i}2 (m + i \gamma^{\perp} \cdot {\cal D}_{\perp})
\int_{-L}^{x^-} dv e^{-ieA_-(x^+)(x^- - v)} \gamma^+ \psi_+(x^+,v,x^{\perp}) .
\label{eq:it2}
\end{eqnarray}
These equations implicitly express $\psi_{\pm}(x^+,x^-,x^{\perp})$ in terms 
of $\psi_+$, for $x^+ = 0$ and $x^- > -L$, and $\psi_-$, for $x^+ > 0$ and 
$x^- = -L$. To make the relation explicit we substitute (\ref{eq:it2}) into 
(\ref{eq:it1}) and iterate.  The result is an infinite series,
\begin{eqnarray}
\lefteqn{\psi_+(x^+,x^-,x^{\perp}) = \sum_{n=0}^{\infty}\left[-\frac{1}{2}(m 
+ i \gamma^{\perp} \cdot {\cal D}_{\perp} ) (m - i \gamma^{\perp} \cdot 
{\cal D}_{\perp} )\right]^n} \nonumber \\
& & \int_{0}^{x^+} du_{1} \int_{-L}^{x^-} dv_1 e^{-ieA_-(u_1) (x^--v_1)} 
\int_0^{u_1} du_2 \int_{-L}^{v_1} dv_{2} e^{-ieA_-(u_2) (v_1-v_2)} 
\nonumber \\
& & \dots \int_{0}^{u_{n-1}} du_n \int_{-L}^{v_{n-1}} dv_n e^{-ieA_-(u_n)
(v_{n-1}-v_n)} \biggl\{\psi_{+}(0,v_n,x^{\perp}) \nonumber \\
& &  - \frac{i}{2} (m + i \gamma^{\perp} \cdot {\cal D}_{\perp})
\int_0^{u_n} du e^{-ieA_-(u_{n}) (v_{n}+L)} \gamma^- \psi_-(u,-L,x^{\perp})
\biggr\}. \label{eq:series}
\end{eqnarray}
This series can be summed as in \cite{TW2}. The result is,
\begin{eqnarray}
\lefteqn{\psi_+(x^+,x^-,x^{\perp}) = \int_{-L}^{\infty}dv \int_{-\infty}^{+
\infty} \frac{dk^+}{2\pi} e^{i(k^++i/L)(v-x^-)}} \nonumber \\
& & \times \biggl\{{\cal U}\left(x^\perp,\tau(0,x^+;k^+)\right)
\psi_+(0,v,x^{\perp}) - \frac{i}{2}(m + i \gamma^{\perp} \cdot {\cal D}_{
\perp}) \nonumber \\
& & \int_0^{x^+} du e^{-ieA_-(u)(v+L)} {\cal U}\left(x^{\perp},\tau(u,x^+;
k^+)\right) \gamma^- \psi_{-}(u,-L,x^{\perp})\biggr\}. \label{series_solution}
\end{eqnarray}
The various hitherto undefined functions are,
\begin{eqnarray}
{\cal U}(x^{\perp},\tau) & \equiv & e^{-i{\cal H}[eA_{\perp}(x_{\perp})]\tau}
\; , \\
{\cal H}[eA_{\perp}(x_{\perp})] & \equiv  & \frac{1}{2}(m + i \gamma^{\perp}
\cdot{\cal D}_{\perp} )(m - i \gamma^{\perp} \cdot {\cal D}_{\perp}) \; ,
\label{eq:series2} \\
\tau(u,x^+;k^+) & \equiv & \int_u^{x^+}\frac{du^{\prime}}{k^+ - e
A_-(u^{\prime}) + i/L} \; .
\end{eqnarray}
To shorten expressions in later sections we make the definitions,
\begin{eqnarray}
\tau_+ \equiv \tau(0,x^+;k^+) & , & \tau_- \equiv \tau(u,x^+;k^+) 
\label{tau1} \\
\tau_+^* \equiv \tau^*(0,x^+;q^+) & , & \tau_-^* \equiv \tau(y,x^+;q^+) 
\label{tau2} \\
\tau_{++} \equiv \tau_+^* - \tau_+ & , & \tau_{--} \equiv \tau_-^* - \tau_- 
\label{tau3}
\end{eqnarray}

At this stage our solution (\ref{series_solution}) is still valid for 
\textit{any} $A_{\perp}(x^{\perp})$, however its dependence upon the
initial value operators is complicated by the transverse covariant 
derivative operator. To exhibit this dependence we express ${\cal U}$
as a kernel,
\be
{\cal U}(x^{\perp},\tau)f(x^{\perp}) \equiv \int d^{2}y^{\perp} 
K(x^{\perp},y^{\perp};\tau)f(y^{\perp}) \; .
\ee
One obtains the kernel by treating ${\cal H}[e A_{\perp}(x^{\perp})]$ as a 
first quantized Hamiltonian. The spinor structure factors out through the 
reduction,
\be
{\cal H}[eA_{\perp}(x^{\perp})] = \frac12 \Bigl[m^{2} - {\cal D}_{\perp}
\cdot {\cal D}_{\perp}\Bigr] + \beta \Sigma^3 \; ,
\ee
where $\Sigma^{3} \equiv \frac{i}2 [\gamma^1,\gamma^2]$ and $\beta \equiv
\frac{\vert e \vert B}2 = - \frac{e B}2$. For our linear $A_{\perp}(x^{\perp})$
(\ref{gaugepots}) the Hamiltonian is that of a rotated, 2-dimensional harmonic 
oscillator. Identifying its kernel is straightforward,
\be
{\cal K}(x^{\perp},y^{\perp};\tau) \equiv e^{-i \beta \Sigma^3 \tau} {\cal G}(
x^{\perp},y^{\perp};\tau) \; .
\ee
The function ${\cal G}(x^{\perp},y^{\perp};\tau)$ is,
\be
- {i \beta e^{-\frac{i}2 m^2 \tau} \over 2 \pi \sin( \beta \tau)} \exp\Biggl[
\frac{i}2 \beta \cot(\beta \tau)(x^{\perp}-y^{\perp})^2 - i e x^{\perp} \cdot 
A_{\perp}(y^{\perp}) \Biggl] \label{kernel} \; .
\ee
We will often use its Fourier transform on $y^{\perp}$,
\begin{eqnarray}
\widetilde{\cal G}(x^{\perp},k^{\perp},s) = \frac{e^{-{i\over2}m^{2}s} 
e^{-ik^{\perp} \cdot x^{\perp}}}{\cos( \beta s)} \exp\left[{i \over 2 \beta} 
\tan( \beta s) (k^{\perp}-eA_{\perp}(x^{\perp}))^{2}\right] \; .
\end{eqnarray}

In terms of the kernel our solution for $\psi_+$ is,
\begin{eqnarray}
\lefteqn{\psi_+(x^+,x^-,x^{\perp}) = \int_{-L}^{\infty} dv \int_{-\infty}^{+
\infty} \frac{dk^+} {2\pi} e^{i(k^++i/L) (v-x^{-})} \int d^{2}y^{\perp}} 
\nonumber \\
& & \times \biggl\{{\cal K}\left(x^\perp,y^{\perp};\tau(0,x^+;k^+)\right)
\psi_+(0,v,y^{\perp}) - \frac{i}{2}(m + i \gamma^{\perp} \cdot {\cal D}_{
\perp}) \nonumber \\
& & \int_{0}^{x^+}due^{-ieA_-(u) (v+L)} {\cal K}\left(x^{\perp},y^{\perp};
\tau(u,x^+;k^+)\right) \gamma^{-} \psi_{-}(u,-L,y^{\perp})\biggr\} \; .
\label{plussolution}
\end{eqnarray}
The solution for $\psi_-$ is obtained by inverting (\ref{em1}). That is,
\be
\psi_- = i\gamma^+(m + i \gamma^{\perp} \cdot {\cal D}_{\perp})^{-1} 
\partial_+ \psi_+ \; . \label{minussolution} 
\ee
The inverse operator can be obtained by noting that $\cal K$ is the Green's 
function for a time dependent Shr\"odinger equation,
\be
\Bigl({\cal H} - i\frac{\partial}{\partial s}\Bigr) {\cal K}(x^{\perp},
y^{\perp};s) = 0 \; . \label{tdse}
\ee
Integrating (\ref{tdse}) gives us the inverse operator,
\be
(m + i \gamma^{\perp} \cdot {\cal D}_{\perp})^{-1} \delta^2(x^{\perp} - 
y^{\perp}) = i \int_0^{\infty} ds \Bigl(m - i \gamma^{\perp} \cdot 
{\cal D}_{\perp}(x_{\perp})\Bigr) {\cal K}(x^{\perp},y^{\perp};s) 
\label{inverse}
\ee

It is sometimes desirable to express dependence upon the transverse 
coordinates using the harmonic oscillator basis of ${\cal H}$. One begins 
by defining first quantized momenta,
\begin{equation}
p_x \equiv -i {\partial \over \partial x} \qquad , \qquad p_y \equiv -i 
{\partial \over \partial y} \; .
\end{equation}
With the position operators these are formed into lowering operators,
\begin{equation}
a_x \equiv {1 \over \sqrt{2 \beta}} \Bigl( \beta x + i p_x \Bigr) \qquad 
, \qquad a_y \equiv {1 \over \sqrt{2 \beta}} \Bigl( \beta y + i p_y 
\Bigr) \; .
\end{equation}
Finally, one defines complex raising and lowering operators,
\begin{equation}
a_{\pm} \equiv {1 \over \sqrt{2}} \Bigl( a_x \pm i a_y \Bigr) \qquad , \qquad
a^{\dagger}_{\pm} \equiv {1 \over \sqrt{2}} \Bigl( a_x^{\dagger} \mp i 
a^{\dagger}_y \Bigr) \; .
\end{equation}
The original coordinates and derivatives have the following expressions,
\begin{eqnarray}
x = {1 \over 2 \sqrt{\beta}} \Bigl( a_+ + a_+^{\dagger} + a_- + 
a_-^{\dagger}\Bigr) & , & y = {i \over 2 \sqrt{\beta}} \Bigl( - a_+ + 
a_+^{\dagger} + a_- - a_-^{\dagger}\Bigr) , \\
\partial_x = {\sqrt{\beta} \over 2} \Bigl( a_+ - a_+^{\dagger} + a_- -
a_-^{\dagger}\Bigr) & , & \partial_y = i {\sqrt{\beta} \over 2} \Bigl( 
- a_+ - a_+^{\dagger} + a_- + a_-^{\dagger}\Bigr) . \qquad
\end{eqnarray}
Hence the covariant derivative operators are,
\begin{eqnarray}
{\cal D}_x & = & \partial_x - i \beta y = \sqrt{\beta} \Bigl( a_- - 
a_-^{\dagger} \Bigr) \; , \\
{\cal D}_y & = & \partial_y + i \beta x = \sqrt{\beta} \Bigl( i a_- +
i a_-^{\dagger} \Bigr) \; .
\end{eqnarray}

The point of this technology is to give a simple expression for the
first quantized Hamiltonian,
\begin{equation}
{\cal H} = \frac12 m^2 + \Bigl( 2 a_-^{\dagger} a_- + 1 + \Sigma^3 \Bigr) 
\beta \; .
\end{equation}
Its normalized eigenstates are,
\begin{equation}
W_{n_{\pm}}(x^{\perp}) \equiv {(a_+^{\dagger})^{n_+} \over \sqrt{n_+ !}}
{(a_-^{\dagger})^{n_-} \over \sqrt{n_- !}} \sqrt{\frac{\beta}{\pi}} 
e^{-\frac{\beta}2 \Vert x^{\perp} \Vert^2} \; .
\end{equation}
The fields can be expressed in this basis as follows,
\begin{equation}
\psi_+(x^+,x^-,n_{\pm}) \equiv \int d^2x^{\perp} W^*_{n_{\pm}}(x^{\perp})
\psi_+(x^+,x^-,x^{\perp}) \equiv \langle \! \langle W_{n_{\pm}} \vert
\psi_+(x^+,x^-) \rangle \! \rangle \; .
\end{equation}
Our solution (\ref{series_solution}) assumes the form,
\begin{equation}
\psi_{\pm}(x^+,x^-,x^{\perp}) = \int_{-\infty}^{\infty} {dk^+ \over 2\pi} 
e^{-i (k^+ + i/L) x^-} \sum_{n_{\pm} = 0}^{\infty} \widetilde{\psi}_{\pm
}(x^+,k^+,n_{\pm}) W_{n_{\pm}}(x^{\perp}) \; ,
\end{equation}
where we define,
\begin{eqnarray}
\lefteqn{\widetilde{\psi}_+(x^+,k^+,n_{\pm}) \equiv \int_{-L}^{\infty} dv
e^{i (k^+ + i/L) v} \Biggl\{ \left\langle \!\! \left\langle W_{n_{\pm}} 
\left\vert e^{-i {\cal H} \tau_+} \right\vert \psi_+(0,v) \right\rangle \!\! 
\right\rangle } \nonumber \\
& & \!\!\! - \frac{i}2 \int_0^{x^+} \!\!\!\!\! du e^{-i e A_-(u) (v + L)} 
\Bigl\langle \!\! \Bigl\langle W_{n_{\pm}} \Bigl\vert (m + i \gamma^{\perp} 
\cdot {\cal D}_{\perp} ) e^{-i {\cal H} \tau_-} \gamma^- \Bigr\vert 
\psi_-(u,-L) \Bigr\rangle \!\! \Bigr\rangle \Biggr\} , \qquad \label{eigen} \\
\lefteqn{\widetilde{\psi}_-(x^+,k^+,n_{\pm}) \equiv \int_{-L}^{\infty} dv
e^{i (k^+ + i/L) v} \Biggl\{ \Bigl\langle \!\! \Bigl\langle W_{n_{\pm}} 
\Bigl\vert \frac{i}{2 {\cal H}} (m + i \gamma^{\perp} \cdot {\cal D}_{\perp} ) 
\gamma^+ } \nonumber \\
& & \!\!\! \times e^{-i {\cal H} \tau_+} \Bigr\vert \psi_+(0,v) \Bigr\rangle 
\!\!  \Bigr\rangle + \int_0^{x^+} \!\!\!\!\! du e^{-i e A_-(u) (v + L)} 
\Bigl\langle \!\! \Bigl\langle W_{n_{\pm}} \Bigl\vert e^{-i {\cal H} \tau_-} 
\Bigr\vert \psi_-(u,-L) \Bigr\rangle \!\! \Bigr\rangle \Biggr\} . \qquad 
\label{eigen-}
\end{eqnarray}
Note that the inverse of ${\cal H}$ is straightforward to evaluate in the
harmonic oscillator basis.

Taking the $x^+$ derivative of (\ref{eigen}) gives,
\begin{eqnarray}
\lefteqn{-i\partial_+ \widetilde{\psi}_+(x^+,k^+,n_{\pm}) = - {\frac12 m^2 +
(2 n_- + 1 + \Sigma^3) \beta \over k^+ - eA_-(x^+) + i/L} \widetilde{\psi}_+(
x^+,k^+,n_{\pm})} \nonumber \\ 
& & - { e^{-i(k^+ + i/L)L} \over k^+ - eA_-(x^+) + i/L} \frac{i}2 \Bigl\langle
\!\! \Bigl\langle W_{n_{\pm}} \Bigl\vert \Bigl(m + i \gamma^{\perp} \cdot 
{\cal D}_{\perp}\Bigr) \gamma^- \Bigr\vert \psi_-(x^+,-L) \Bigr\rangle \!\!
\Bigr\rangle .
\end{eqnarray}
The last term only contributes at $k^+ = eA_-(x^+)$ in the large L limit 
since,
\be
\lim_{L \rightarrow \infty} \frac{ e^{-i(k^+ - eA_-(x^+) + i/L)L}}{k^+ -
eA_-(x^+) + i/L} = -2\pi i\delta\left(k^+ - eA_-(x^+)\right).
\ee
We see that the large $L$ limit of $\widetilde{\psi}_+(x^+,k^+,n_{\pm})$ is 
an eigen-operator of $-i \partial_+$! Since the eigenvalues of $\Sigma^3$ are
$\pm 1$ the sign of the eigenvalue is controlled by the denominator $k^+ = e 
A_-(x^+)$. For $k^+ > e A_-(x^+)$ the large $L$ limit of $\widetilde{\psi}_+(
x^+,k^+,n_{\pm})$ annihilates electrons with spin $s = \frac12 \Sigma^3$; for 
$k^+ < e A_-(x^+)$ it creates positrons with spin $s = -\frac12 \Sigma^3$.

\section{Light-cone quantization}

The Lagrangian density for Dirac fermions in our background is,
\begin{eqnarray}
\lefteqn{{\cal L} = \sqrt{2} \psi_+^{\dagger} \Bigl(i\partial_+ \psi_+
- \frac12 (m + i \gamma^{\perp} \cdot {\cal D}_{\perp}) \gamma^- \psi_-\Bigr)} 
\nonumber \\
& & + \sqrt{2} \psi_{-}^{\dagger} \Bigl((i\partial_{-} - eA_{-}) \psi_- - 
\frac12 (m + i \gamma^{\perp} \cdot {\cal D}_{\perp}) \gamma^+ \psi_+\Bigr).
\label{lagrangian}
\end{eqnarray}
Using (\ref{lagrangian}) we may read off the algebra that our operator 
solutions satisfy on the initial value surfaces. The conjugate momenta of 
these initial value fields, $\psi_+(0,v,x^{\perp})$ and $\psi_-(u,-L,x^{
\perp})$, are the normal derivatives of (\ref{lagrangian}) evaluated on the 
surfaces $x^+ = 0$ and $x^- = -L$, respectively. Therefore, the  momentum 
conjugate to $\psi_+(0,v,x^{\perp})$ is $i\sqrt{2} \psi_+^{\dagger}(0,v,x^{
\perp})$, and the corresponding conjugate momentum to $\psi_-^{\dagger
}(u,-L,x^{\perp})$ is $i\sqrt{2} \psi_-(u,-L,x^{\perp})$. The two initial 
value surfaces are spacelike separated, and therefore the two nonzero 
anti-commutators are,
\begin{eqnarray}
\Bigl\{ \psi_+(0,v,x^{\perp}) , \psi_+^{\dagger}(0,w,y^{\perp}) \Bigr\} & = & 
\frac1{\sqrt{2}} P_+ \delta (v-w) \delta^2(x^{\perp}-y^{\perp}), 
\label{ac1} \\
\Bigl\{ \psi_-(u,-L,x^{\perp}) , \psi_-^{\dagger}(y,-L,y^{\perp}) \Bigr\} 
& = & \frac1{\sqrt{2}} P_- \delta(u-y) \delta^2(x^{\perp}-y^{\perp}). 
\label{ac2}
\end{eqnarray}
The anti-commutation relations for arbitrary equal $x^+$ and equal $x^-$ 
are not independent but follow from our solutions 
(\ref{plussolution},\ref{minussolution}), 
\begin{eqnarray}
\Bigl\{ \psi_+(x^+,x^-,x^{\perp}) , \psi_+^{\dagger}(x^+,y^-,y^{\perp})
\Bigr\} & = & \frac1{\sqrt{2}} P_+ \delta (x^- - y^-) \delta^2(x^{\perp}
- y^{\perp}), \; \\
\Bigl\{ \psi_-(x^+,x^-,x^{\perp}) , \psi_-^{\dagger}(y^+,x^-,y^{\perp})
\Bigr\} & = & \frac1{\sqrt{2}} P_- \delta(x^+ - y^+) \delta^2(x^{\perp}
- y^{\perp}). \; 
\end{eqnarray}

It remains to specify the Heisenberg state. For our purposes the natural
``vacuum'' $|\Omega\rangle$ is empty at $x^+ = 0$ and $x^- = -L$. This makes 
calculating expectation values of fermion bilinears straightforward. One 
first uses our solution to express the bilinear in terms of the initial 
value operators, and then computes the expectation value of these in the 
absence of the background fields using the standard free vacuum,
\begin{eqnarray}
\lefteqn{ \langle \Omega | \psi_{\alpha}(x^+,x^-,x^{\perp}) \psi_{\beta}^{
\dagger}(y^+,y^-,y^{\perp}) | \Omega\rangle_{A_{\mu}=0}} \nonumber \\
& & = \int {d^3p \over (2\pi)^3} {(\sla{p} \gamma^{0} + m \gamma^0)_{\alpha
\beta} \over 2\omega} e^{-ip^- (x^+ - y^+) - ip^+ (x^- - y^-) +i p^{\perp}
\cdot (x^{\perp} - y^{\perp})} , \label{vev1}
\end{eqnarray}
\begin{eqnarray}
\lefteqn{ \langle \Omega | \psi_{\beta}^{\dagger}(y^+,y^-,y^{\perp})
\psi_{\alpha}(x^+,x^-,x^{\perp}) | \Omega\rangle_{A_{\mu}=0}} \nonumber \\
& & = \int {d^3p \over (2\pi)^3} {(\sla{p} \gamma^0 - m \gamma^0)_{\alpha
\beta} \over 2\omega} e^{ip^- (x^+ - y^+) + ip^+ (x^- - y^-) - i p^{\perp}
\cdot (x^{\perp} - y^{\perp})} . \label{vev2}
\end{eqnarray}
The variable of integration above is $p^i$ and we define $\omega \equiv 
\sqrt{m^2 + \vec{p} \cdot \vec{p}}$.

In using (\ref{vev1}-\ref{vev2}) one first specializes to the desired initial
value position and spinor component. Next change variables from $p^3$ to
either $p^+$ or $p^-$,
\be
\int_{-\infty}^{\infty} \!\!\!\!\!\!\! dp^3 = \int_0^{\infty} \!\!\!\!\!\! 
dp^+ {\omega \over p^+} = \int_0^{\infty} \!\!\!\!\!\! dp^- {\omega \over 
p^-} \; .
\ee
The complementary light-cone momentum is given by the mass shell condition, 
$2 p^+ p^- = m^2 + p^{\perp} \cdot p^{\perp}$.

When the spinor indices are not explicitly written we shall understand 
expectation values of the form $\psi^{\dagger} M \psi$ to involve an implied 
spinor trace. Specializing to the initial value surfaces and taking $\pm$ 
components gives the various combinations of this form,
\be
\langle \Omega | \psi_+^{\dagger}(0,w,y^{\perp}) \psi_+(0,v,x^{\perp}) | 
\Omega\rangle = \sqrt{2} \delta^2(x^{\perp} - y^{\perp}) \int_0^{\infty}
\!\! {dp^+ \over 2\pi} e^{ip^+ (v-w)}, \label{lcvev1} 
\ee
\begin{eqnarray}
\lefteqn{ \langle \Omega | \psi_+^{\dagger}(0,w,y^{\perp}) \gamma^-
\psi_-(u,-L,x^{\perp}) | \Omega\rangle} \nonumber \\
& & \hspace{1.5cm} = - \sqrt{2} \int {d^2p^{\perp} \over (2\pi)^2} e^{-i 
p^{\perp} \cdot (x^{\perp} - y^{\perp})} \int_0^{\infty} \!\! {dp^+ \over 
2\pi} {m \over p^+} e^{ip^- u - ip^+ (w+L)}, \label{lcvev2}
\end{eqnarray}
\begin{eqnarray}
\lefteqn{\langle \Omega | \psi_-^{\dagger}(y,-L,y^{\perp}) \gamma^+
\psi_+(0,v,x^{\perp}) | \Omega\rangle} \nonumber \\
& & \hspace{1.5cm} = - \sqrt{2} \int {d^2p^{\perp} \over (2\pi)^2} e^{-i 
p^{\perp} \cdot (x^{\perp} - y^{\perp})} \int_0^{\infty} \!\! {dp^+ \over 
2\pi} {m \over p^+} e^{-ip^- y + ip^+ (v+L)}, \label{lcvev3} \\
\lefteqn{ \langle \Omega | \psi_-^{\dagger}(y,-L,y^{\perp}) \psi_-(u,-L,x^{
\perp}) | \Omega\rangle} \nonumber \\
& & \hspace{3cm} = \sqrt{2} \delta^2(x^{\perp} - y^{\perp}) \int_0^{\infty} 
\!\! {dp^- \over 2\pi} e^{ip^- (u-y)} , \\ \label{lcvev4}
& & = \sqrt{2} \delta^2(x^{\perp} - y^{\perp}) {1 \over 2}\left\{\delta(u-y)
+ {i \over \pi} {\cal{P}} \left({1 \over u-y}\right)\right\} .
\end{eqnarray}
Notice that in the $L\rightarrow \infty$ limit (\ref{lcvev2},\ref{lcvev3}) 
vanish. This means that the transverse coordinate dependence only contributes
delta functions!

In addition to (\ref{lcvev1}-\ref{lcvev4}), more complicated spinor traces 
will appear. It is convenient to list two of the operator reductions here to 
expedite derivations in later sections,
\begin{eqnarray}
\lefteqn{ \langle \Omega | \psi^{\dagger}_+(0,w,y^{\perp\prime}) e^{ i \beta
\Sigma^3 \tau_{\pm\pm}} \psi_+(0,v,x^{\perp\prime}) | \Omega \rangle} 
\nonumber \\
& & \ = \sqrt{2} \delta^2(x^{\perp\prime} - y^{\perp\prime}) \cos( \beta
\tau_{\pm\pm}) \int_0^{\infty} \!\! {dp^+ \over 2\pi} e^{ip^+ (v-w)}, 
\label{lcvev5} \\ 
\lefteqn{ \langle \Omega | \psi^{\dagger}_-(y,-L,y^{\perp\prime}) \gamma^+
e^{i \beta \Sigma^3 \tau^{\ast}_{\pm}} \left(m + i \gamma^{\perp} \cdot 
{\cal D}_{\perp}^{\ast}(x^{\perp}) \right) {\cal{G}}^*(x^{\perp},y^{\perp
\prime};\tau^{\ast}_{\pm})} \nonumber \\
& & \times \left(m + i \gamma^{\perp} \cdot {\cal D}_{\perp}(x^{\perp}) 
\right) {\cal{G}}^{\ast}(x^{\perp},x^{\perp\prime};\tau_{\pm})
e^{-i \beta \Sigma^3 \tau_{\pm}} \gamma^- \psi_-(u,-L,x^{\perp\prime})
| \Omega\rangle \nonumber \\
& & = \sqrt{8} \delta^2(x^{\perp\prime} - y^{\perp\prime}) {\cal{G}}^{\ast}(
x^{\perp},y^{\perp\prime};\tau^{\ast}_{\pm}) \biggl\{ \left(m^2 + 
\overleftarrow{\cal D}_{\perp}^{\ast} \cdot \overrightarrow{\cal D}_{\perp}
\right) \cos(\beta \tau_{\pm\pm}) \nonumber \\
& & - \epsilon^{ij} \overleftarrow{{\cal{D}}}_{i\perp}^{\ast} 
\overrightarrow{{\cal{D}}}_{j\perp} \sin(\beta \tau_{\pm\pm}) \biggl\}
{\cal{G}}(x^{\perp},x^{\perp\prime};\tau_{\pm}) \int_0^{\infty} \!\! {dp^- 
\over 2\pi} e^{ip^{-}(u-y)} \label{lcvev6},
\end{eqnarray}
where $\epsilon^{ij}$ is the anti-symmetric Levi-Civita density in two
dimensions with $i$ and $j$ running over the values 1 and 2.

\section{Pair creation probability}

At the end of Section 2 we were able to identify an operator $\widetilde{
\psi}_+(x^+,k^+,n_{\pm})$ which gives exact eigenstates of the light-cone 
evolution operator $-i \partial_+$ in the large $L$ limit. Its behavior 
changes abruptly at time $x^+ = X(k^+)$, defined such that $k^+ \equiv e 
A_-\Bigl(X(k^+)\Bigr)$. For $x^+ < X(k^+)$ the operator $\widetilde{\psi}_+(
x^+,k^+,n_{\pm})$ annihilates electrons of momentum $k^+$, Landau level 
$n_-$ and spin $\frac12 \Sigma^3$. For $x^+ > X(k^+)$ it creates positrons 
of momentum $k^+$, Landau level $n_-$ and spin $-\frac12 \Sigma^3$. 

The transition between these two regimes is a manifestation of particle 
creation, which is an instantaneous event on the light-cone. Just as in 
the previous treatments \cite{TW1,TW2}, the newly created positron accelerates 
to the speed of light in the $+x^3$ direction, so its worldline is 
asymptotically parallel to the $x^+$ axis. The electron goes the other way, 
so its worldline is asymptotically parallel to the $x^-$ axis. This has a 
curious effect when one regards $x^+$ as the evolution operator: electrons 
leave the light-cone manifold while the positrons accumulate. 

In this section we compute the probability ${\rm Prob}(k^+,n_-,s)$ for creating 
a positron of momentum $k^+$, Landau level $n_-$ and spin $s$. From the 
previous section we see that the two nonzero spinor components of $\widetilde{
\psi}_+(x^+,k^+,n_{\pm})$ lack only a factor of $2^{\frac14}$ to be 
canonically normalized. Therefore we can extract the creation probability (for 
$x^+ > X(k^+)$ from the relation,
\begin{eqnarray}
\lefteqn{ \lim_{L \rightarrow \infty} \sqrt{2} \left\langle \Omega \left\vert
\widetilde{\psi}^{\dagger}_+(x^+,q^+,m_{\pm}) ({\scriptstyle \frac12} - s
\Sigma^3) \widetilde{\psi}_+(x^+,k^+,n_{\pm}) \right\vert \Omega \right\rangle}
\nonumber \\
& & \hspace{2cm} = \left[1 - {\rm{Prob}}(k^+,n_-,s) \right] 2\pi \delta(k^+ - 
q^+) \delta_{m_{\pm} , n_{\pm}} . \label{probability}
\end{eqnarray}

The procedure for evaluating (\ref{probability}) is to first express the
operators in terms of the initial value operators using 
(\ref{eigen}-\ref{eigen-}). For any bilinear this produces four kinds of 
operator products: the $++$ combination in which each term is from the $x^+ 
= 0$ surface; the $+-$ combination in which the first is from $x^+ = 0$ and 
the second from $x^- = -L$; and so on. We then compute the expectation values 
of each product from the free, sourceless theory as explained in the last 
section. Finally, the large $L$ limit is taken. Since causality permits only 
the $++$ and $--$ products to survive this limit, we report only these terms.

The $++$ product is simple,
\begin{eqnarray}
\lefteqn{ \sqrt{2} \left\langle \Omega \left\vert \widetilde{\psi}^{\dagger
}_+(x^+,q^+,m_{\pm}) ({\scriptstyle \frac12} - s \Sigma^3) \widetilde{\psi
}_+(x^+,k^+,n_{\pm}) \right\vert \Omega \right\rangle_{++}} \nonumber \\
& & = \int_{-L}^{+\infty} dv e^{i(k^+ + i/L)v} \int_{-L}^{+\infty} dw
e^{-i(q^+ - i/L)w} \nonumber \\
& & \times \Bigl\langle \!\! \Bigl\langle W_{n_{\pm}} \Bigl\vert {\rm Tr}\left[
e^{-i {\cal H} \tau_+} \Bigl({\scriptstyle \frac12} - s \Sigma^3\Bigr) P_+ 
\int_0^{\infty} {dp^+ \over 2\pi} e^{i p^+ (v - w)} e^{i {\cal H} \tau_+^*}
\right] \Bigr\vert W_{m_{\pm}} \Bigr\rangle \!\! \Bigr\rangle , \qquad \\
& & = \int_0^{\infty} {dp^+ \over 2\pi} \, {e^{-i (k^+ + p^+ + i/L) L} \over
k^+ + p^+ + i/L} \, {e^{i (q^+ + p^+ - i/L) L} \over q^+ + p^+ - i/L}
\delta_{m_{\pm} , n_{\pm}} e^{i \epsilon(n_-,s) \tau_{++}} \; , \label{++int}
\end{eqnarray}
where $\epsilon(n_-,s) \equiv \frac12 m^2 + (2 n_- + 1 - 2 s) \beta$. We are 
interested in the limit $L \rightarrow \infty$, in which case,
\begin{equation}
{e^{-i (k^+ + p^+ + i/L) L} \over k^+ + p^+ + i/L} \, {e^{i (q^+ + p^+ - i/L) 
L} \over q^+ + p^+ - i/L} \longrightarrow 2\pi \delta(k^+ + p^+) \, 2\pi
\delta(q^+ + p^+) \; .
\end{equation}
When $q^+ = k^+$ the large $L$ limit of expression (\ref{tau3}) for $\tau_{++}$
becomes, 
\begin{eqnarray}
\lefteqn{\lim_{L \rightarrow \infty} \int_0^{x^+} \left({du \over k^+ - 
eA_-(u) - i/L} - {du \over k^+ - eA_-(u) + i/L} \right) } \nonumber \\
& & \hspace{4cm} = 2 {\pi} i X^{\prime}(k^+) \theta(eA_- - k^+) \theta(k^+) .
\label{taupluslimit}
\end{eqnarray}
This vanishes in (\ref{++int}) because the delta functions and the range of
$p^+$ conspire to make $k^+$ negative whereas $e A_-(x^+)$ is assumed 
positive, 
\begin{eqnarray}
\lefteqn{ \lim_{L \rightarrow \infty} \sqrt{2} \left\langle \Omega \left\vert 
\widetilde{\psi}^{\dagger}_+(x^+,q^+,m_{\pm}) ({\scriptstyle \frac12} - s 
\Sigma^3) \widetilde{\psi}_+(x^+,k^+,n_{\pm}) \right\vert \Omega 
\right\rangle_{++}} \nonumber \\
& & \hspace{5cm} = 2\pi \delta(k^+ - q^+) \theta(- k^+) \delta_{m_{\pm} , 
n_{\pm}} \; . \label{plusprobability}
\end{eqnarray}

The $--$ term is a little more difficult, 
\begin{eqnarray}
\lefteqn{ \sqrt{2} \left\langle \Omega \left\vert \widetilde{\psi}^{\dagger
}_+(x^+,q^+,m_{\pm}) ({\scriptstyle \frac12} - s \Sigma^3) \widetilde{\psi
}_+(x^+,k^+,n_{\pm}) \right\vert \Omega \right\rangle_{--}} \nonumber \\
& & = \frac14 \int_{-L}^{+\infty} \!\!\!\!\!\!\! dv e^{i(k^+ + i/L)v} 
\int_0^{x^+} \!\!\!\!\!\!\! du e^{-i e A_-(u) (v + L)} 
\int_{-L}^{+\infty} \!\!\!\!\!\!\! dw e^{-i(q^+ - i/L)w} 
\int_0^{x^+} \!\!\!\!\!\!\! dy e^{i e A_-(y) (w + L)} \nonumber \\
& & \times \Bigl\langle \!\! \Bigl\langle \!\!\! \Bigl\langle W_{n_{\pm}} 
\Bigl\vert {\rm Tr}\Biggl[ \Bigl(m + i \gamma^{\perp} \cdot {\cal D}_{\perp} 
\Bigr) e^{-i {\cal H} \tau_-} \Bigl({\scriptstyle \frac12} - s \Sigma^3\Bigr) 
\gamma^- P_- \nonumber \\
& & \hspace{2cm} \times \int_0^{\infty} {dp^- \over 2\pi} e^{-i p^- (u - y)} 
\gamma^+ e^{i {\cal H} \tau_-^*} \Bigl(m - i \gamma^{\perp} \cdot 
{\cal D}_{\perp} \Bigr) \Biggr] \Bigr\vert W_{m_{\pm}} \Bigr\rangle \!\! 
\Bigr\rangle \; , \\
& & = \frac12 \int_0^{x^+} \!\!\!\!\!\!\! du {e^{-i (k^+ + i/L) L} \over
k^+ - e A_-(u) + i/L} \int_0^{x^+} \!\!\!\!\!\!\! dy {e^{i (q^+ - i/L) L} 
\over q^+ - e A_-(y) - i/L} \nonumber \\
& & \hspace{1cm} \times \left\{ \frac12 \delta(u - y) + {i \over 2 \pi} 
P\left({1 \over u - y} \right) \right\} \Bigl\langle \!\! \Bigl\langle \!\!\! 
\Bigl\langle W_{n_{\pm}} \Bigl\vert {\rm Tr}\Biggl[ \Bigl(m + i \gamma^{\perp} 
\cdot {\cal D}_{\perp} \Bigr) \nonumber \\
& & \hspace{3cm} \times P_+ \Bigl({\scriptstyle \frac12} - s \Sigma^3\Bigr) 
e^{i {\cal H} \tau_{--} } \Bigl(m - i \gamma^{\perp} \cdot {\cal D}_{\perp} 
\Bigr) \Biggr] \Bigr\vert W_{m_{\pm}} \Bigr\rangle \!\!  \Bigr\rangle \; , 
\end{eqnarray}
We reduce the transverse structure using the identity,
\begin{eqnarray}
\lefteqn{\Bigl(m + i \gamma^{\perp} \cdot {\cal D}_{\perp} \Bigr) P_+ \Bigl(
{\scriptstyle \frac12} - s \Sigma^3\Bigr) e^{i {\cal H} \tau_{--}} \Bigl(m - 
i \gamma^{\perp} \cdot {\cal D}_{\perp} \Bigr) } \nonumber \\
& & \hspace{4cm} = P_+ \Bigl( {\scriptstyle \frac12} - s \Sigma^3\Bigr) 
2 {\cal H} e^{i {\cal H} \tau_{--}} \; .
\end{eqnarray}
This brings the $--$ term to the interesting form,
\begin{eqnarray}
\lefteqn{ \sqrt{2} \left\langle \Omega \left\vert \widetilde{\psi}^{\dagger
}_+(x^+,q^+,m_{\pm}) ({\scriptstyle \frac12} - s \Sigma^3) \widetilde{\psi
}_+(x^+,k^+,n_{\pm}) \right\vert \Omega \right\rangle_{--}} \nonumber \\
& & = \int_0^{x^+} \!\!\!\!\!\!\! du {e^{-i (k^+ + i/L) L} \over
k^+ - e A_-(u) + i/L} \int_0^{x^+} \!\!\!\!\!\!\! dy {e^{i (q^+ - i/L) L} 
\over q^+ - e A_-(y) - i/L} \nonumber \\
& & \hspace{1cm} \times \left\{ \frac12 \delta(u - y) + {i \over 2 \pi} 
P\left({1 \over u - y} \right) \right\} \delta_{m_{\pm} , n_{\pm}} 
\epsilon(n_-,s) e^{i \epsilon(n_-,s) \tau_{--}} \; . \qquad \label{aha!}
\end{eqnarray}

At this stage we observe that (\ref{aha!}) is the same as the $1+1$ expression
(4.8) of ref. \cite{TW2} with the trivial replacement,
\begin{equation}
\frac12 m^2 \longrightarrow \frac12 m^2 + (2 n_- + 1 - 2s) \beta \equiv
\epsilon(n_-,s) \; . \label{eps}
\end{equation}
This means that the remaining analysis has already been done! We can read 
the final result off from ref. \cite{TW2},
\begin{eqnarray}
\lefteqn{ \lim_{L \rightarrow \infty} \sqrt{2} \left\langle \Omega \left\vert 
\widetilde{\psi}^{\dagger}_+(x^+,q^+,m_{\pm}) ({\scriptstyle \frac12} - s 
\Sigma^3) \widetilde{\psi}_+(x^+,k^+,n_{\pm}) \right\vert \Omega 
\right\rangle_{--}} \nonumber \\
& & = 2\pi \delta(k^+ - q^+) \delta_{m_{\pm} , n_{\pm}} \theta(k^+) 
\theta\Bigl(e A_-(x^+) - k^+\Bigr) \Bigl[ 1 - e^{-2 \pi \lambda(k^+,n_-,s)}
\Bigr] \; , \qquad \label{minusprobability} \end{eqnarray}
where we define,
\begin{equation}
\lambda(k^+,n_-,s) \equiv {\epsilon(n_-,s) \over \vert e E(X(k^+)) \vert} \; .
\label{lambda} \end{equation}
Note that we could have used this same procedure to shorten the $++$ 
derivation as well. It is almost always the case that expressing the 
transverse coordinate dependence in the harmonic oscillator basis results
in an expression which differs only by the replacement (\ref{eps}) from one
already computed in ref. \cite{TW2} for the $1+1$ dimensional theory.

Combining (\ref{plusprobability}) and (\ref{minusprobability}) gives,
\begin{eqnarray}
\lefteqn{ \lim_{L \rightarrow \infty} \sqrt{2} \left\langle \Omega \left\vert 
\widetilde{\psi}^{\dagger}_+(x^+,q^+,m_{\pm}) ({\scriptstyle \frac12} - s 
\Sigma^3) \widetilde{\psi}_+(x^+,k^+,n_{\pm}) \right\vert \Omega 
\right\rangle} \nonumber \\
& & = 2 \pi \delta(k^+ - q^+) \delta_{m_{\pm} , n_{\pm}} \Biggl\{ \theta(-k^+)
\nonumber \\
& & \hspace{3cm}+ \theta(k^+) \theta\Bigl(eA_-(x^+) - k^+\Bigr) \Bigl[1 - 
e^{-2 \pi \lambda(k^+,n_-,s)}\Bigr] \Biggr\} \; .
\end{eqnarray}
The $\theta(-k^+)$ term implies there is no particle creation for $k^+ < 0$.
These modes start out as positron creation operators and they continue to have
that meaning for $E(x^+) > 0$. Since the state was initially empty of these
modes it remains so. The other term implies that positrons are created for
$0 < k^+ < e A_-(x^+)$ with probability,
\begin{equation}
{\textnormal{Prob}}(k^+,n_-,s) = e^{-2 \pi \lambda(k^+,n_-,s)} \; . 
\label{prob}
\end{equation}
Note that the spin dependence makes physical sense. It is more probable for
a positron to be created with its spin aligned ($s = + \frac12$) with the 
magnetic field than opposed ($s = -\frac12$).

\section{The vector currents}

Our operator solutions (\ref{plussolution},\ref{minussolution}) enable us to 
calculate exactly the one-loop response to an external electromagnetic field. 
The light-cone currents are,
\be
J^{\pm} = {e \over \sqrt{2}} \left(\psi_{\pm}^{\dagger} \psi_{\pm} - 
{\rm{Tr}}[ \psi_{\pm} \psi_{\pm}^{\dagger}] \right).
\ee
As usual in quantum field theory, we must regulate these operators. We 
accomplish this by point splitting $J^{\pm}$ in $x^{\perp}$ and in $x^{\mp}$. 
To maintain gauge invariance we add a gauge string when needed,
\begin{eqnarray}
\lefteqn{ J^+(x^+;x^-,y^-;x^{\perp},y^{\perp}) = {e \over \sqrt{2}} e^{i e
A_-(y^- - x^-) +i e y^{\perp} \cdot A_{\perp}(x^{\perp})} \biggl(\psi^{
\dagger}_+(x^+,x^-,x^{\perp}) } \nonumber \\*
& & \hspace{2cm} \times \psi_+(x^+,y^-,y^{\perp}) - {\rm{Tr}}\left[ 
\psi_+(x^+,y^-,y^{\perp}) \psi_+^{\dagger}(x^+,x^-,x^{\perp}) \right]\biggl) . 
\quad \label{jplussplit} \\
\lefteqn{ J^-(x^+,y^+;x^-;x^{\perp},y^{\perp}) = {e \over \sqrt{2}} e^{i e
y^{\perp} \cdot A_{\perp}(x^{\perp})} \biggl( \psi^{\dagger}_-(x^+,x^-,x^{
\perp}) } \nonumber \\*
& & \hspace{2cm} \times \psi_-(y^+,x^-,y^{\perp}) - {\rm{Tr}}\left[ 
\psi_-(y^+,x^-,y^{\perp}) \psi_-^{\dagger}(x^+,x^-,x^{\perp}) \right]\biggl) . 
\quad \label{jminussplit}
\end{eqnarray}
Point splitting breaks Hermiticity. Therefore, our currents are the symmetric
limits of (\ref{jplussplit},\ref{jminussplit}),
\begin{equation}
J^{\pm}(x) = \lim_{y \rightarrow x} {1 \over 2} \left(J^{\pm}(x;y) + 
J^{\pm}(y;x) \right). \quad \label{jsymmetric} 
\end{equation}
Note that the expectation values of the transverse currents $J^{1},\ J^{2}$ 
vanish. This can be seen by simple Dirac algebra. It corresponds physically
to the zero average transverse current for a particle undergoing helical 
motion.

We begin with $J^+$ and compute the expectation values of the $++$ and 
$--$ terms as in the previous section. (As before, the $+-$ and $-+$
terms vanish in the large $L$ limit.) The reductions are similar to those 
in the previous section, so we show the results (with $y^- = x^- + \Delta^-$)
after performing the $v$ and $w$ integrations and taking the transverse 
expectation values,
\begin{eqnarray}
\lefteqn{\Bigl\langle \Omega \Bigl\vert J^+(x^+;x^-,y^-;x^{\perp},y^{\perp})
\Bigr\vert \Omega \Bigr\rangle_{++} = \sum_{n_{\pm} , s} W^*_{n_{\pm}}(x^{
\perp}) W_{n_{\pm}}(y^{\perp}) e^{ie y^{\perp} \cdot A_{\perp}(x^{\perp})} }
\nonumber \\
& & \times \frac{e}2 e^{i e A_-(x^+) \Delta^-} \Bigl\{ \int_{-\infty}^0 - 
\int_0^{\infty} \Bigr\} {dp^+ \over 2 \pi} \int_{-\infty}^{\infty} {dk^+ \over 
2\pi} {e^{-i (k^+ + i/L) (y^- + L)} \over k^+ - p^+ + \frac{i}{L}} \nonumber\\
& & \hspace{3cm} \times \int_{-\infty}^{\infty} {dq^+ \over 2 \pi} {e^{+i (q^+ 
- i/L) (x^- + L)} \over q^+ - p^+ - \frac{i}{L}} e^{i \epsilon(n_-,s) 
\tau_{++}} . \qquad \\
\lefteqn{\Bigl\langle \Omega \Bigl\vert J^+(x^+;x^-,y^-;x^{\perp},y^{\perp})
\Bigr\vert \Omega \Bigr\rangle_{--} = \sum_{n_{\pm} , s} W^*_{n_{\pm}}(x^{
\perp}) W_{n_{\pm}}(y^{\perp}) e^{i e y^{\perp} \cdot A_{\perp}(x^{\perp})} }
\nonumber \\
& & \times \frac{e}2 e^{i e A_-(x^+) \Delta^-} \int_0^{x^+} \!\!\!\!\! du 
\int_0^{x^+} \!\!\!\!\! dy \Bigl\{\int_{-\infty}^0 - \int_0^{\infty}
\Bigr\} {dp^- \over 2 \pi} e^{-i p^- (u - y)} \int_{-\infty}^{\infty} {dk^+ 
\over 2\pi} \nonumber \\
& & \times {e^{-i (k^+ + i/L) (y^- + L)} \over k^+ - e A_-(u) + \frac{i}{L}}
\int_{-\infty}^{\infty} {dq^+ \over 2\pi} {e^{+i (q^+ - i/L) (x^- + L)} \over 
q^+ - e A_-(y) - \frac{i}{L}} \epsilon(n_-,s) e^{i \epsilon(n_-,s) 
\tau_{--}} . \qquad
\end{eqnarray}
Recall that $\tau_{++}$ and $\tau_{--}$ were defined in (\ref{tau1}-\ref{tau3})
and that $\epsilon(n_-,s) \equiv \frac12 m^2 + (2 n_- + 1 - 2 s) \beta$. 

Each of these results has the form of $\sum_{n_{\pm} , s} W^*_{n_{\pm}}(x^{
\perp}) W_{n_{\pm}}(y^{\perp}) e^{i e y^{\perp} \cdot A_{\perp}(x^{\perp})}$ 
times the corresponding $1+1$ dimensional result of \cite{TW2} with the 
trivial replacement: $\frac12 m^2 \longrightarrow \epsilon(n_-,s)$. We can 
therefore read off the large $L$ limits directly. That for the $++$ terms 
follows from equation (5.10) of \cite{TW2},
\begin{eqnarray}
\lefteqn{\lim_{L \rightarrow \infty} \Bigl\langle \Omega \Bigl\vert 
J^+(x^+;x^-,y^-;x^{\perp},y^{\perp}) \Bigr\vert \Omega \Bigr\rangle_{++} = 
\sum_{n_{\pm} , s} W^*_{n_{\pm}}(x^{\perp}) W_{n_{\pm}}(y^{\perp}) e^{i e 
y^{\perp} \cdot A_{\perp}(x^{\perp})} } \nonumber \\
& & \hspace{1cm} \times \frac{e}2 \left\{ {i \over \pi \Delta^-} - \int_0^{e 
A_-} {dp^+ \over 2\pi} \left[1 + e^{-2 \pi \lambda(p^+,n_-,s)} \right] e^{-i 
(p^+ - e A_-) \Delta^-} \right\} . \quad 
\end{eqnarray}
The large $L$ limit of the $--$ terms derives from equations (5.11-5.13) of 
\cite{TW2},
\vfill\eject

\begin{eqnarray}
\lefteqn{\lim_{L \rightarrow \infty} \Bigl\langle \Omega \Bigl\vert 
J^+(x^+;x^-,y^-;x^{\perp},y^{\perp}) \Bigr\vert \Omega \Bigr\rangle_{--} = 
\sum_{n_{\pm} , s} W^*_{n_{\pm}}(x^{\perp}) W_{n_{\pm}}(y^{\perp}) e^{i e 
y^{\perp} \cdot A_{\perp}(x^{\perp})} } \nonumber \\
& & \hspace{2.5cm} \times \frac{e}2 \int_0^{e A_-} {dp^+ \over 2\pi} \left[1 
- e^{-2 \pi \lambda(p^+,n_-,s)} \right] e^{-i (p^+ - e A_-) \Delta^-} 
\; . \qquad
\end{eqnarray}

Combining the $++$ and $--$ terms gives,
\begin{eqnarray}
\lefteqn{\lim_{L \rightarrow \infty} \Bigl\langle \Omega \Bigl\vert 
J^+(x^+;x^-,y^-;x^{\perp},y^{\perp}) \Bigr\vert \Omega \Bigr\rangle = \sum_{
n_{\pm} , s} W^*_{n_{\pm}}(x^{\perp}) W_{n_{\pm}}(y^{\perp}) e^{i e y^{\perp}
\cdot A_{\perp}(x^{\perp})} } \nonumber \\
& & \hspace{2cm} \times e \left\{ {i \over 2 \pi \Delta^-} - \int_0^{e A_-} 
{dp^+ \over 2\pi} e^{-2 \pi \lambda(p^+,n_-,s)} e^{-i (p^+ - e A_-) \Delta^-} 
\right\} \; . \qquad
\end{eqnarray}
At this stage we can take $y^{\perp} \rightarrow x^{\perp}$. Hermitization
discards the $1/\Delta^-$ term, at which point we can also take $\Delta^- 
\rightarrow 0$. The result is,
\begin{equation}
\lim_{L \rightarrow \infty} \Bigl\langle \Omega \Bigl\vert J^+(x^+,x^-,x^{
\perp}) \Bigr\vert \Omega \Bigr\rangle = - e \sum_{n_{\pm} , s} \Vert 
W_{n_{\pm}}(x^{\perp}) \Vert^2 
\int_0^{e A_-} {dp^+ \over 2\pi} e^{-2 \pi \lambda(p^+,n_-,s)} \; .
\end{equation}
This expression has a transparent physical interpretation based on the role
of $J^+$ as the light-cone charge density. This charge density derives from 
the steady accumulation of positrons as the electron member of each newly
created pair leaves the light-cone manifold. Hence the charge density is the 
sum over states of the $-e$ contributed by each positron, times the pair 
production probability (\ref{prob}) we derived in Section 4.

Since the expectation value of the current operators cannot depend upon the
transverse coordinate we may as well set $x^{\perp} = 0$. The harmonic 
oscillator basis functions are especially simple at this point,
\begin{equation}
W_{n_{\pm}}(0) = (-)^{n_-} \delta_{n_- , n_+} \sqrt{\beta \over \pi} \; .
\end{equation}
Recalling that $\lambda(p^+,n_-,s) = \epsilon(n_-,s)/\vert e E(X(p^+)) \vert$,
we can perform the sums over $n_{\pm}$ and $s$,
\begin{eqnarray}
\lim_{L \rightarrow \infty} \Bigl\langle \Omega \Bigl\vert J^+(x^+,x^-,x^{
\perp}) \Bigr\vert \Omega \Bigr\rangle \!\!\!\! & = & \!\!\!\! -{e \beta 
\over 2 \pi^2} \int_0^{e A_-} \!\!\!\!\!\!\! dp^+ e^{-\frac{\pi m^2}{ \vert 
e E \vert}} \Bigl[1 \! + \! e^{-\frac{4 \pi \beta}{\vert e E \vert}} \Bigr] 
\sum_{n = 0}^{\infty} \Bigl( e^{-\frac{4 \pi \beta}{\vert e E \vert}}\Bigr)^n 
\!\! , \quad \\
\!\!\!\! & = & \!\!\!\! {e^2 B \over 4 \pi^2} \int_0^{e A_-} \!\!\!\!\!\!\!
dp^+ e^{-\frac{\pi m^2}{ \vert e E \vert}} {\rm coth}\Bigl[{\scriptstyle 
\frac{\pi B}{E(X(p^+))}}\Bigr] \; . \label{J+final}
\end{eqnarray}
Since $\beta \equiv -\frac{eB}2$ we see that the $B \longrightarrow 0$ limit
agrees with the result of \cite{TW1}. The other new limit, that of large $B$,
seems more interesting. In that case the hyperbolic cotangent goes to one,
so $J^+$ grows linearly in the magnetic field. This might be phenomenologically
relevant to astrophysics because very large, approximately homogeneous magnetic
fields are known to occur. For example, the magnetic field strength in a 
neutron star can reach $B \sim 10^{13}~{\rm Gauss}$ over a kilometer coherence 
length.

We turn now to $J^-$. After performing the $v$ and $w$ integrations and taking 
the transverse expectation values the $++$ and $--$ terms assume the form,
\begin{eqnarray}
\lefteqn{\Bigl\langle \Omega \Bigl\vert J^-(x^+,y^+;x^-;x^{\perp},y^{\perp})
\Bigr\vert \Omega \Bigr\rangle_{++} = \sum_{n_{\pm} , s} W^*_{n_{\pm}}(x^{
\perp}) W_{n_{\pm}}(y^{\perp}) e^{ie y^{\perp} \cdot A_{\perp}(x^{\perp})} }
\nonumber \\
& & \times \frac{e}2 {\partial \over \partial x^+} {\partial \over \partial 
y^+} \Bigl\{ \int_{-\infty}^0 - \int_0^{\infty} \Bigr\} {dp^+ \over 2 \pi} 
\int_{-\infty}^{\infty} {dk^+ \over 2\pi} {e^{-i (k^+ + i/L) (x^- + L)} \over 
k^+ - p^+ + \frac{i}{L}} \nonumber \\
& & \hspace{3cm} \times \int_{-\infty}^{\infty} {dq^+ \over 2\pi} {e^{+i (q^+ - 
i/L) (x^- + L)} \over q^+ - p^+ - \frac{i}{L}} {e^{i \epsilon(n_-,s) \sigma_+} 
\over \epsilon(n_-,s)} . \qquad \label{ud++} \\
\lefteqn{\Bigl\langle \Omega \Bigl\vert J^-(x^+,y^+;x^-;x^{\perp},y^{\perp})
\Bigr\vert \Omega \Bigr\rangle_{--} = \sum_{n_{\pm} , s} W^*_{n_{\pm}}(x^{
\perp}) W_{n_{\pm}}(y^{\perp}) e^{i e y^{\perp} \cdot A_{\perp}(x^{\perp})} }
\nonumber \\
& & \times \frac{e}2 {\partial \over \partial x^+} {\partial \over \partial 
y^+} \int_0^{ y^+} \!\!\!\!\! du \int_0^{x^+} \!\!\!\!\! dy \Bigl\{\int_{-
\infty}^0 - \int_0^{\infty} \Bigr\} {dp^- \over 2 \pi} e^{-i p^- (u - y)} 
\int_{-\infty}^{\infty} {dk^+ \over 2\pi} \nonumber \\
& & \times {e^{-i (k^+ + i/L) (x^- + L)} \over k^+ - e A_-(u) + \frac{i}{L}}
\int_{-\infty}^{\infty} {dq^+ \over 2\pi} {e^{+i (q^+ - i/L) (x^- + L)} \over 
q^+ - e A_-(y) - \frac{i}{L}} e^{i \epsilon(n_-,s) \sigma_-} . \label{ud--}
\end{eqnarray}
The quantities $\sigma_{\pm}$ are just $\tau_{++}$ and $\tau_{--}$ with the
upper limits of the second integral in each changed from $x^+$ to $y^+$,
\begin{eqnarray}
\sigma_+ & \equiv & \int_0^{x^+} {du' \over q^+ - e A_-(u') - \frac{i}{L}}
- \int_0^{y^+} {du' \over k^+ - e A_-(u') + \frac{i}{L}} \; , \\
\sigma_- & \equiv & \int_y^{x^+} {du' \over q^+ - e A_-(u') - \frac{i}{L}}
- \int_u^{y^+} {du' \over k^+ - e A_-(u') + \frac{i}{L}} \; .
\end{eqnarray}

The $x^-$ derivative of $J^-$ is ultraviolet finite so the oscillator
sums again multiply the same $1+1$ dimensional currents whose large $L$
limits were already computed in \cite{TW2}. For example, the large $L$ limit
of the $++$ terms follows from equations (5.23-5.24) of \cite{TW2},
\begin{eqnarray}
\lefteqn{\lim_{L \rightarrow \infty} \partial_- \Bigl\langle \Omega \Bigl\vert 
J^-(x^+,x^+;x^-;x^{\perp},y^{\perp}) \Bigr\vert \Omega \Bigr\rangle_{++} }
\nonumber \\
& & = \sum_{n_{\pm} , s} W^*_{n_{\pm}}(x^{\perp}) W_{n_{\pm}}(y^{\perp}) 
e^{i e y^{\perp} \cdot A_{\perp}(x^{\perp})} {e^2 E(x^+) \over 4 \pi} 
\left[1 - e^{-2 \pi \lambda(p^+,n_-,s)} \right] \; . \qquad
\end{eqnarray}
The large $L$ limit of the $--$ terms derives from equations (5.25-5.28) of 
\cite{TW2},
\begin{eqnarray}
\lefteqn{\lim_{L \rightarrow \infty} \partial_- \Bigl\langle \Omega \Bigl\vert 
J^-(x^+,x^+;x^-;x^{\perp},y^{\perp}) \Bigr\vert \Omega \Bigr\rangle_{--} }
\nonumber \\
& & = \sum_{n_{\pm} , s} W^*_{n_{\pm}}(x^{\perp}) W_{n_{\pm}}(y^{\perp}) 
e^{i e y^{\perp} \cdot A_{\perp}(x^{\perp})} {e^2 E(x^+) \over 4 \pi} 
\left[-1 - e^{-2 \pi \lambda(p^+,n_-,s)} \right] \; . \qquad
\end{eqnarray}
Once the two terms are combined we can take the transverse coordinates to
coincidence and again exploit transverse translational invariance to perform 
the sums over $n_{\pm}$ and $s$ at $x^{\perp} = 0$,
\begin{eqnarray}
\lim_{L \rightarrow \infty} \partial_- \Bigl\langle \Omega \Bigl\vert J^-(x^+,
x^-,x^{\perp}) \Bigr\vert \Omega \Bigr\rangle \!\!\! & = & \!\!\!\!\! - {e^2 
E(x^+) \over 2 \pi} \sum_{n_{\pm} , s} \Vert W_{n_{\pm}} \Vert^2 
e^{-2 \pi \lambda(e A_-,n_-,s)} \; , \qquad \\
\!\!\! & = & \!\!\!\! {e^3 E(x^+) B \over 4 \pi^2} e^{-\frac{\pi m^2}{ \vert 
e E \vert}} {\rm coth}\Bigl[{\scriptstyle \frac{\pi B}{E(x^+)}}\Bigr] \; .
\label{dJ-final}
\end{eqnarray}
Comparison with (\ref{J+final}) verifies current conservation.

We can obtain the undifferentiated current $J^-$ by integrating with respect 
to $x^-$, just as in $1+1$ dimensions \cite{TW2}. However, the $3+1$ 
dimensional integration constant must be treated with care. Although our 
choice of state makes the expectation value of $J^-(x^+,-L,x^{\perp})$ vanish,
moving even infinitesimally to the left of $x^- = -L$ results in an ultraviolet
divergence! Of course this is the one loop photon field strength 
renormalization. To extract it we fix one of the fields at $x^- = -L$ and take 
the other just inside. Since there is no $++$ term, and the mixed terms always 
vanish for large $L$, we compute only the $--$ contribution,
\begin{eqnarray}
\lefteqn{{e \over \sqrt{2}} e^{i e A_-(x^+) \Delta^- + i e A_{\perp}(x^{\perp})
\cdot \Delta^{\perp}} \Bigl\langle \Omega \Bigl\vert \Bigl\{ \psi_-^{\dagger
}(x^+,-L,x^{\perp}) \psi_-(x^+,\Delta^- - L,x^{\perp} + \Delta^{\perp}) } 
\nonumber \\
& & \hspace{1.5cm} - {\rm Tr}\Bigl[\psi_-(x^+,\Delta^- - L,x^{\perp} + 
\Delta^{\perp}) \psi_-^{\dagger}(x^+,-L,x^{\perp})\Bigr] \Bigr\} \Bigr\vert 
\Omega \Bigr\rangle_{--} = \nonumber \\
& & \sum_{n_{\pm} , s} W^*_{n_{\pm}}(x^{\perp}) W_{n_{\pm}}(x^{\perp} \! + \!
\Delta^{\perp}) e^{i e A_{\perp}(x^{\perp}) \cdot \Delta^{\perp}} 
\frac{e}{4\pi} \int_{-\infty}^{\infty} \!\!\!\!\! dk^+ {e^{-i (k^+ - e A_-(x^+)
+ \frac{i}{L}) \Delta^- } \over k^+ - e A_-(x^+) + \frac{i}{L}} \nonumber \\
& & \times \!\! \int_0^{x^+} \!\!\!\!\! {du \over k^+ - e A_-(u) + \frac{i}{L}} 
\Bigl\{\int_{-\infty}^0 \!\!\! - \int_0^{\infty} \Bigr\} {dp^- \over 2 \pi} 
e^{i p^- (x^+ - u)} \epsilon(n_-,s) e^{-i \epsilon(n_-,s) \tau_-} . \quad
\label{UV}
\end{eqnarray}
Because of its significance to this analysis we remind the reader of the 
function $\tau_- = \tau(u,x^+;k^+)$ from (\ref{tau1}),
\begin{equation}
\tau(u,x^+;k^+) = \int_u^{x^+} {du' \over k^+ - e A_-(u') + \frac{i}{L}} \; .
\end{equation}
It will be important to note that $\tau(u,x^+;k^+)$ has a negative
imaginary part.

The next step is to perform the oscillator and spin sums using the relation,
\begin{eqnarray}
\lefteqn{\sum_{n_{\pm} , s} W^*_{n_{\pm}}(x^{\perp}) W_{n_{\pm}}(y^{\perp}) 
e^{i e A_{\perp}(x^{\perp}) \cdot \Delta^{\perp}} \epsilon(n_-,s) 
e^{-i \epsilon(n_-,s) \tau_-} } \nonumber \\
& & \hspace{2cm} = i {\partial \over \partial \tau_-} \Biggl\{- {i \beta \over 
\pi} e^{-\frac{i}2 m^2 \tau_-} \cot(\beta \tau_-) e^{\frac{i}2 \beta 
\cot(\beta \tau_-) \Vert \Delta^{\perp} \Vert^2} \Biggr\} \; .
\end{eqnarray}
Since $x^+ \geq u$ the $p^-$ integral gives,
\begin{equation}
\Bigl\{\int_{-\infty}^0 \!\!\! - \int_0^{\infty} \Bigr\} {dp^- \over 2 \pi} 
e^{i p^- (x^+ - u)} = - \frac{i}{2 \pi} {1 \over x^+ - u} \; .
\end{equation}
Now change variables from $u$ to $\tau(u,x^+;k^+)$ by recognizing the complex
differential,
\begin{equation}
d\tau = {\partial \tau \over \partial u} du = {-du \over k^+ - e A_-(u) +
\frac{i}{L}} \; .
\end{equation}
Since $\tau(0,x^+;k^+) \equiv \tau_+$ and $\tau(x^+,x^+;k^+) = 0$, expression
(\ref{UV}) takes the form,
\begin{eqnarray}
\lefteqn{- \frac{i e}{8 \pi^3} \int_{-\infty}^{\infty} \!\!\!\!\! dk^+ {e^{-i 
[k^+ - e A_-(x^+) + \frac{i}{L}] \Delta^- } \over k^+ - e A_-(x^+) + 
\frac{i}{L}} \int_0^{\tau_+} {d\tau \over x^+ - u} } \nonumber \\
& & \hspace{3cm} \times {\partial \over \partial \tau} \Bigl\{ \beta 
\cot(\beta \tau) e^{-\frac{i}2 m^2 \tau + \frac{i}2 \beta \cot(\beta \tau) 
\Vert \Delta^{\perp} \Vert^2} \Bigr\} \; . 
\end{eqnarray}
Note that the negative imaginary part of $\tau$ makes the integrand 
exponentially suppressed as $\tau \longrightarrow 0$ as long as $\Vert 
\Delta^{\perp} \Vert^2 \neq 0$.

We must next express $1/(x^+ - u)$ in terms of $\tau$. First expand 
$\tau(u,x^+;k^+)$ for small $\Delta u \equiv x^+ - u$,
\begin{eqnarray}
\lefteqn{\tau(u,x^+;k^+) = {\Delta u \over k^+ - e A_-(x^+) + \frac{i}{L}} + 
{\frac12 e A_-'(x^+) {\Delta u}^2 \over [k^+ - e A_-(x^+) + \frac{i}{L}]^2} }
\nonumber \\
& & + {\frac16 e A_-''(x^+) {\Delta u}^3 \over [k^+ - e A_-(x^+) + 
\frac{i}{L}]^2} + {\frac13 [e A_-'(x^+)]^2 {\Delta u}^3 \over [k^+ - e A_-(x^+) 
+ \frac{i}{L}]^3} + O({\Delta u}^4) . \qquad 
\end{eqnarray}
Since all the vector potentials are evaluated at $x^+$ we can suppress their 
arguments in subsequent expressions. We also define the complex parameter
$K \equiv k^+ - e A_- + \frac{i}{L}$. Solving perturbatively for $1/\Delta u$
gives, 
\begin{equation}
{1 \over \Delta u} = {1 \over K \tau} \Bigl\{1 + \frac12 e A_-' \tau
+\frac16 e A_-'' K \tau^2 + \frac1{12} (e A_-' \tau)^2 + O(\tau^3) \Bigr\} \; .
\end{equation}
Substituting this result and integrating by parts brings (\ref{UV}) to the 
form,
\begin{eqnarray}
\lefteqn{- \frac{i e}{8 \pi^3} \int_{-\infty}^{\infty} \!\!\!\!\! dk^+ {e^{-i K
\Delta^- } \over K} \Biggl\{ {\beta \over x^+} \cot(\beta \tau_+) e^{-\frac{i}2 
m^2 \tau_+ + \frac{i}2 \beta \cot(\beta \tau_+) \Vert \Delta^{\perp} \Vert^2} }
\nonumber \\
& & \hspace{2cm} + \int_0^{\tau_+} d\tau {\beta \over K \tau^2} \cot(\beta 
\tau) e^{-\frac{i}2 m^2 \tau + \frac{i}2 \beta \cot(\beta \tau) \Vert 
\Delta^{\perp} \Vert^2} \nonumber \\
& & \hspace{3cm} \times \Bigl[1 - \frac16 e A_-'' K \tau^2 - \frac1{12} 
(e A_-' \tau)^2 + O(\tau^3) \Bigr] \Biggr\} \; . 
\end{eqnarray}
Note that the surface term is obviously finite in the unregulated limit.

The ultraviolet divergence derives from the integration over small $\tau$.
From the expansion $\cot(x) = \frac1{x} - \frac13 x - \frac1{45} x^3 + \dots$ 
we infer,
\begin{eqnarray}
\lefteqn{{\beta \over \tau^2} \cot(\beta \tau) e^{-\frac{i}2 m^2 \tau + 
\frac{i}2 \beta \cot(\beta \tau) \Vert \Delta^{\perp} \Vert^2} = e^{\frac{i}{
2\tau} \Vert \Delta^{\perp} \Vert^2} \Biggl\{ {1 \over \tau^3}
- \frac{i}2 \Bigl[m^2 + \frac13 \beta^2 \Vert \Delta^{\perp} \Vert^2\Bigr]
{1 \over \tau^2} } \nonumber \\
& & \hspace{1cm} - \Bigl[\frac18 m^4 + \frac1{12} m^2 \beta^2 \Vert 
\Delta^{\perp} \Vert^2 + \frac1{72} \beta^4 \Vert \Delta^{\perp} \Vert^4 + 
\frac13 \beta^2\Bigr] {1 \over \tau} + O(1) \Biggr\} . \qquad
\end{eqnarray}
Only the following integrals can produce divergences,
\begin{eqnarray}
\int_0^{\tau_+} {d\tau \over \tau^3} e^{\frac{i}{2 \tau} \Delta^2}
& = & \Bigl[-{4 \over \Delta^4} + {2 i \over \tau_+ \Delta^2} \Bigr] 
e^{\frac{i}{2 \tau+} \Delta^2} = -{4 \over \Delta^4} + O(1) \; , \\
\int_0^{\tau_+} {d\tau \over \tau^2} e^{\frac{i}{2 \tau} \Delta^2} 
& = & {2 i \over \Delta^2} e^{\frac{i}{2 \tau+} \Delta^2} = {2 i \over 
\Delta^2} + O(1) \; , \\
\int_0^{\tau_+} {d\tau \over \tau} e^{\frac{i}{2 \tau} \Delta^2} & = & 
-{\rm Ei}\Bigl({i \Delta^2 \over 2 \tau_+}\Bigr) = - \ln(\Delta^2 )
+ O(1) \; .
\end{eqnarray}
We can therefore isolate the terms from (\ref{UV}) which diverge with
the transverse point splitting $\Delta^2 \equiv \Vert \Delta^{\perp} \Vert^2$,
\begin{equation}
\frac{-i e}{8 \pi^3} \int_{-\infty}^{\infty} \!\!\!\!\! dk^+ {e^{-i K \Delta^-} 
\over K^2} \Biggl\{- {4 \over \Delta^4} + {m^2 \over \Delta^2} + \Bigl[{m^2
\over 8} + {\beta^2 \over 3} + {e A_-'' K \over 6} + {(e A_-')^2 \over 12}
\Bigr] \ln(\Delta^2) + O(1) \Biggr\} .
\end{equation}

It remains to perform the $k^+$ integration. Most of the transverse divergences
are proportional to $1/K^2$ so they vanish with $\Delta^-$,
\begin{equation}
\int_{-\infty}^{\infty} \!\!\!\!\! dk^+ {e^{-i K \Delta^-} \over K^2} =
-2 \pi \Delta^- \; .
\end{equation}
They are also purely imaginary and would vanish upon Hermitization. The single
exception is the term proportional to $e A_-''$. The $k^+$ integral for it is,
\begin{equation}
\int_{-\infty}^{\infty} \!\!\!\!\! dk^+ {e^{-i K \Delta^-} \over K} =
-2 \pi i \; .
\end{equation}
This gives a real term which survives when $\Delta^- \longrightarrow 0$,
\begin{eqnarray}
\lefteqn{\lim_{\Delta^- \rightarrow 0} {e \over \sqrt{2}} e^{i e A_- \Delta^- 
+ i e A_{\perp} \cdot \Delta^{\perp}} \Bigl\langle \Omega \Bigl\vert \Bigl\{ 
\psi_-^{\dagger}(x^+,-L,x^{\perp}) \psi_-(x^+,\Delta^- \!-\! L,x^{\perp} \!+\! 
\Delta^{\perp}) } \nonumber \\
& & \hspace{2cm} - {\rm Tr}\Bigl[\psi_-(x^+,\Delta^- - L,x^{\perp} + \Delta^{
\perp}) \psi_-^{\dagger}(x^+,-L,x^{\perp})\Bigr] \Bigr\} \Bigr\vert \Omega 
\Bigr\rangle \nonumber \\
& & = {e^2 A_-''(x^+) \over 24 \pi^2} \ln\Bigl(\Vert \Delta^{\perp} \Vert^{-2}
\Bigr) + {\rm finite} \; , \\
& & = - \delta Z_3 \partial_{\nu} F^{\nu -} + {\rm finite} \; .
\end{eqnarray}
So we have recovered the standard one loop result for the photon field strength
renormalization \cite{BD}. This is another impressive check on the correctness
and consistency of the formalism. As one might expect, the divergence can be
isolated without taking the large $L$ limit.

\section{The axial vector anomaly}

The vector currents we have just obtained give the exact one-loop response to 
our electromagnetic background. Since they are not entire functions of the 
electric field they could never be obtained in a perturbative expansion. It
seems obvious that we can also access some of the nonperturbative structure
of the axial vector currents. This is interesting because it allows one to
check for nonperturbative corrections to the axial vector anomaly, just as 
has already been done in $1+1$ dimensions \cite{TW2}.

The axial vector anomaly is the violation of the naive divergence equation,
\be
\partial_{\mu}J^{\mu}_{5} - 2imJ_{5} = 0 \; .
\ee
The anomaly in electrodynamics results from the one loop triangle diagram 
containing two vector and one pseudo-vector vertices. Adler and Bardeen 
showed that this diagram receives no {\it perturbative} corrections \cite{AB}. 
However, the possibility for nonperturbative corrections remains open. 

Modulo operator ordering and regularization, the axial vector current operator
and its pseudoscalar partner are,
\begin{eqnarray}
J_5^{\pm}& = & \sqrt{2} \psi^{\dagger}_{\pm} \gamma_5 \psi_{\pm} , \\
J_{5} & = & \frac{1}{\sqrt{2}} \left( \psi^{\dagger}_+ \gamma^- \gamma_5
\psi_- + \psi_-^{\dagger} \gamma^+ \gamma_5 \psi_+ \right) .
\end{eqnarray}
The conventions of Section 2 imply, $\gamma_5 \equiv \left(\begin{array}{r r} 
-1 & 0 \\ 0 & 1 \end{array}\right)$. We regulate the axial currents the same
as we did the vector currents,
\begin{eqnarray}
J^+_5(x^+;x^-,y^-,x^{\perp}) & \equiv & \frac{1}{\sqrt{2}} e^{ieA_-(x^+)
\Delta_-} \Biggl\{\psi^{\dagger}_+(x^+,y^-,x^{\perp}) \gamma_5 
\psi_+(x^+,x^-,x^{\perp}) \nonumber \\
& & \hspace{1cm} - {\rm Tr}\biggl[ \gamma_5 \psi_+(x^+,x^-,x^{\perp})
\psi^{\dagger}_+(x^+,y^-,x^{\perp}) \biggl] \Biggl\} . \qquad \\
J^-_5(x^+,y+;x^-,x^{\perp}) & \equiv & \frac{1}{\sqrt{2}} \Biggl\{ \psi^{
\dagger}_+(y^+,x^-,x^{\perp}) \gamma_5 \psi_+(x^+,x^-,x^{\perp}) \nonumber \\
& & \hspace{1cm} - {\rm Tr}\biggl[ \gamma_5 \psi_+(x^+,x^-,x^{\perp})
\psi^{\dagger}_+(y^+,x^-,x^{\perp}) \biggl]\Biggl\} . \qquad 
\end{eqnarray}
The pseudo-scalar is regulated by point splitting in both null directions,
\begin{eqnarray}
\lefteqn{J_5(x^+,y^+;x^-,y^-,x^{\perp}) \equiv \frac{1}{\sqrt{8}} \exp{\left[ 
i e(x^- - y^-) \int_0^{1} \!\! d\eta A_-\left(y^+ + \eta(x^+ - y^+) 
\right)\right]}} \nonumber \\
& & \times \Biggl\{ \psi_+^{\dagger}(y^+,y^-,x^{\perp}) \gamma^- \gamma_5 
\psi_-(x^+,x^-,x^{\perp}) \nonumber \\
& & \hspace{1.5cm} + \psi_-^{\dagger}(y^+,y^-,x^{\perp}) \gamma^+ \gamma_5 
\psi_+(x^+,x^-,x^{\perp}) \nonumber \\
& & \hspace{3cm} - {\rm Tr}\left[ \gamma^- \gamma_5 \psi_-(x^+,x^-,x^{\perp})
\psi_+^{\dagger}(y^+,y^-,x^{\perp})\right] \nonumber \\
& & \hspace{4cm} - {\rm Tr}\left[\gamma^+ \gamma_5 \psi_+(x^+,x^-,x^{\perp})
\psi_-^{\dagger}(y^+,y^-,x^{\perp})\right]\Biggl\} . \qquad \label{j5}
\end{eqnarray}
We Hermitize these operators as we did for the vector current,
\begin{eqnarray}
J^+_5(x^+,x^-,x^{\perp}) \!\! & \equiv & \!\! \lim_{y^{-}\to x^{-}} \frac12 
\left\{J^+_5(x^+;x^-,y^-,x^{\perp}) + J^+_5(x^+;y^-,x^-,x^{\perp}) \right\} ,
\qquad \\
J^-_5(x^+,x^-x^{\perp}) \!\! & \equiv & \!\! \lim_{y^+ \to x^+} \frac12 \left\{
J^-_5(x^+,y^+;x^-,x^{\perp}) + J^-_5(y^+,x^+;x^-,x^{\perp})\right\} . \qquad
\end{eqnarray}

As with the vector currents the subscripts $\pm\pm$ denote which of the four 
initial value products is being considered. Also as before, only the $++$
and $--$ products contribute to the large $L$ limit. We begin with $\langle 
\Omega |J^{+}_{5}|\Omega \rangle$. The $++$ and $--$ expectation values are,
\begin{eqnarray}
\lefteqn{ \langle\Omega | J^+_5(x^+;x^-,y^-,x^{\perp}) | \Omega\rangle_{++}
= \left( \int_{-\infty}^0 \!\!\! - \!\! \int_0^{\infty} \right) {dp^+ \over 
2\pi} e^{-i[p^+ - eA_-(x^+)] \Delta_-} } \nonumber \\*
& & \times \int_{-\infty}^{\infty} {dk^+ \over 2\pi} \frac{e^{-i (k^+ + i/L)
(y^- + L)}} {k^+ - p^+ + i/L} \int_{-\infty}^{\infty} {dq^+ \over 2\pi}
\frac{e^{i (q^+ - i/L) (x^- + L)}}{q^+ - p^+ - i/L} \nonumber \\
& & \times \int d^2x^{\perp \prime} {\cal G}(x^{\perp},x^{\perp \prime};\tau_+)
{\cal G}^{\ast}(x^{\perp},x^{\perp \prime};\tau^{\ast}_+) i \sin(\beta
\tau_{++}), \\
\lefteqn{ \langle \Omega | J^+_5(x^+;x^-,y^-,x^{\perp}) | \Omega\rangle_{--}
= \frac{i}{2} e^{ie A_- \Delta_-} \int_0^{x^+} \!\!\!\! du \int_0^{x^+} 
\!\!\!\! dy {i \over \pi} {\cal{P}}\left({1 \over u-y}\right) } \nonumber \\
& & \times \int_{-\infty}^{\infty} {dk^+ \over 2\pi} \frac{e^{-i (k^+ + i/L)
(y^- + L)}}{k^+ - eA_-(u) + i/L} \int_{-\infty}^{\infty} {dq^+ \over 2\pi}
\frac{e^{i(q^+ - i/L) (x^- + L)}}{q^+ - eA_-(y) - i/L} \nonumber \\
& & \times \int d^2x^{\perp \prime} {\cal{G}}^{\ast}(x^{\perp},x^{\perp
\prime};\tau^{\ast}_-) \Biggl\{\left(m^2 - \overleftarrow{{ \cal{D}}^{\ast}_{
\perp}} \cdot \overrightarrow{{\cal{D}}_{\perp}} \right) \sin(\beta\tau_{--}) 
\nonumber \\
& & - \epsilon^{ij} \overleftarrow{{ \cal{D}}}_{i\perp}^* \overrightarrow{{ 
\cal{D}}}_{j\perp} \cos(\beta \tau_{--}) \Biggl\} {\cal{G}}(x^{\perp},x^{\perp 
\prime};\tau_-) .
\end{eqnarray}

The presence of $\gamma_5$ has interchanged the sines and cosines from where 
they would have resided had we computed the analogous vector current in 
transverse coordinate space. This small change allows us to obtain the result
to all orders without going to the harmonic oscillator basis. For example, the
$++$ term is,
\begin{eqnarray}
\lefteqn{ \langle\Omega | J^+_5(x^+;x^-,y^-,x^{\perp}) | \Omega\rangle_{++}
= {eB \over 4\pi} \left(\int_{-\infty}^0 \!\!\! - \!\! \int_0^{+\infty} 
\right) {dp^+ \over 2\pi} e^{-i [p^+ - eA_-(x^+)] \Delta_-} } \nonumber \\*
& & \times \int_{-\infty}^{\infty} {dk^+ \over 2\pi} \frac{e^{-i (k^+ + i/L)
(y^- + L)}}{k^+ - p^+ + i/L} \int_{-\infty}^{\infty} {dq^+ \over 2\pi}
\frac{e^{i (q^+ - i/L) (x^- + L)}}{q^+ - p^+ - i/L} e^{{i \over 2} m^2
\tau_{++}} , \\
& & \longrightarrow {eB \over 4\pi} \left(\int_{-\infty}^0 \!\!\! - \!\!
\int_0^{+\infty} \right) {dp^+ \over 2\pi} e^{-i [p^+ - eA_-] \Delta_-} 
e^{-2 \pi \lambda(p^+) \theta(p^+) \theta(eA_- - p^+)} \nonumber \\
& & = {eB \over 4\pi} \Biggl\{{i \over \pi \Delta_-} - \int_0^{eA_-} \! {dp^+
\over 2\pi} \left[1 + e^{-2 \pi \lambda(p^+)} \right] e^{-i (p^+ - eA_-)
\Delta_-}\Biggl\}, \label{j5plusplus}
\end{eqnarray}
where $\lambda(p^+) \equiv \lambda(p^+,0,\frac12)$, and $\lambda(p^+,n_-,s)$
was defined in (\ref{lambda}).

The $--$ term can be greatly simplified by the identity,
\begin{eqnarray}
\lefteqn{ \left({\cal D}^*_{\perp} {\cal G}^*(x^{\perp},x^{\perp \prime};
\tau^*_-) \right) \cdot \left({\cal D}_{\perp} {\cal G}(x^{\perp},x^{\perp 
\prime};\tau_-) \right) \sin(\beta \tau_{--})} \nonumber \\
& & \!\!\! +\epsilon^{ij} \left({\cal D}^*_i {\cal G}^*(x^{\perp},x^{\perp 
\prime};\tau^*_-) \right) \left({\cal D}_j {\cal G}(x^{\perp},x^{\perp 
\prime};\tau_-) \right) \cos(\beta \tau_{--}) = 0.
\end{eqnarray}
Using this identity and taking the large $L$ limit gives,
\begin{eqnarray}
\lefteqn{ \langle\Omega | J^+_5(x^+;x^-,y^-,x^{\perp}) | \Omega\rangle_{--}
= {im^2 \over 2} e^{ieA_- \Delta_-} \int_0^{x^+} \!\!\!\! du \int_0^{x^+} 
\!\!\!\!  dy {\cal P} \left({1 \over u-y} \right) } \nonumber \\
& & \times \int_{- \infty}^{\infty} {dk^+ \over 2\pi} \frac{e^{-i (k^+ + i/L)
(y^- + L)}}{k^+ - eA_-(u) + i/L} \int_{-\infty}^{\infty} {dq^+ \over 2\pi}
\frac{e^{i (q^+ - i/L) (x^- + L)}}{q^+ - eA_-(y) - i/L} \nonumber \\
& & \times \int d^2x^{\perp \prime} \sin(\beta \tau_{--}) {\cal G}(x^{\perp},
x^{\perp \prime};\tau_-) {\cal G}^*(x^{ \perp},x^{\perp \prime},\tau^*_-) , \\
& & \longrightarrow {eB \over 4 \pi} \int_0^{eA_-} \!\!\!\!\! dp^+ 
\lambda(p^+) e^{-i(p^+ - eA_-) \Delta_-} e^{-2 \pi \lambda(p^+)} \nonumber \\
& & \qquad\times \int_{-\infty}^{\infty} {da \over 2\pi} \frac{e^{-i (a+i)}}{
a+i} e^{-i \lambda(p^+) \ln(a+i)} \int_{-\infty}^{\infty} {db \over 2\pi}
\frac{e^{i (b-i)}}{b-i} e^{i \lambda(p^+) \ln(b-i)} , \qquad \\
& & = {eB \over 4\pi} \int_0^{eA_-} \!\!\!\!\! dp^+ \lambda(p^+) e^{-i(p^+ - 
eA_-(x^+)) \Delta_-} \left[ \frac{1 - e^{-2 \pi \lambda(p^+)}}{2 \pi \lambda(
p^+)}\right] \; , \\
& & = {eB \over 8 \pi^2} \int_0^{eA_-} \!\!\!\!\! dp^+ \left[1 - e^{-2 \pi 
\lambda(p^+)} \right] \; . \label{j5plusminus}
\end{eqnarray}
Combining (\ref{j5plusplus}) and (\ref{j5plusminus}) and Hermitizing gives 
us the large $L$ expression for $J^+_{5}$,
\begin{equation}
\lim_{L \rightarrow \infty} \langle \Omega | J^+_5(x^+,x^-,x^{\perp}) | 
\Omega\rangle = -{eB \over 4\pi^2} \int_0^{eA_-} \!\!\!\!\! dp^+ e^{-2 \pi
\lambda(p^+)} \; .
\end{equation}

$J^-_5$ involves many of the same procedures. Beginning with the $++$ term, 
it has the following reduction,
\begin{eqnarray}
\lefteqn{ \langle \Omega | J^-_{5}(x^+,y^+;x^-,x^{\perp}) | \Omega\rangle_{++}
= } \nonumber \\*
& &- {im^2 \over 2} \int_{-\infty}^{\infty} {dk^+ \over 2\pi} \frac{e^{-i (k^+
+ i/L) (x^- + L)}}{k^+ - eA_-(x^+) + i/L} \int_{-\infty}^{\infty} {dq^+
\over 2\pi} \frac{e^{i (q^+ - i/L) (x^- + L)}}{q^+ - eA_-(y^+) - i/L} 
\nonumber \\
& & \times \left( \int_{-\infty}^0 \!\!\! - \!\! \int_0^{\infty} \right) 
{dp^+ \over 2\pi} \frac1{k^+ - p^+ + i/L} \frac1{q^+ - p^+ - i/L} \nonumber \\
& & \times\int d^2x^{\perp\prime} \sin(\beta \tau_{++}) {\cal{G}}(x^{
\perp},x^{\perp \prime},\tau_+) {\cal{G}}^{\ast}(x^{\perp},x^{\perp 
\prime},\tau^{\ast}_+) \; , \\
& & \!\!\!\! = -{eB m^2 \over 8\pi} \int_{-\infty}^{\infty} {dk^+ \over 2\pi}
\frac{e^{-i (k^+ + i/L) (x^- + L)}}{k^+ - eA_-(x^+) + i/L} \int_{-\infty}^{
\infty} {dq^+ \over 2\pi} \frac{e^{i(q^+ - i/L) (x^- + L)}}{q^+ - eA_-(y^+)
- i/L} \nonumber \\
& & \qquad \times \frac1{k^+ - q^+ + 2i/L} \Biggl[-i + {1 \over \pi} \ln\left(
{k^+ + i/L \over q^+ - i/L} \right) \Biggl] e^{{i \over 2} m^2\tau_{++}} \; .
\label{lastex}
\end{eqnarray}
We again take the $x^-$ derivative to complete the calculation, this time 
requiring the axial currents to vanish at $x^- = -L$.\footnote{That this is
so can be seen in (\ref{lastex}) from the fact that the $k^+$ and $q^+$ 
integrals can be closed above and below to avoid each's respective poles.}
Acting $\partial_-$ on the $++$ term, taking the large $L$ limit, and 
enforcing coincidence gives,
\begin{eqnarray}
\lim_{L \rightarrow \infty} \partial_- \langle \Omega \vert J^-_{5}(x^+,x^-,
x^{\perp}) \vert \Omega \rangle_{++} & = & {eB m^2 \over 8\pi} \int_{-\infty
}^{\infty} {da \over 2\pi} \frac{e^{-i (a+i)}}{a+i} e^{i \lambda(eA_-) 
\ln(a+i)} \nonumber \\
& & \hspace{.5cm} \times \int_{-\infty}^{\infty} {db \over 2\pi} 
\frac{e^{i(b-i)}}{b-i} e^{-i \lambda(eA_-) \ln(b-i)} , \qquad \\
& = & -{e^2 E(x^+) B \over 8\pi^2} \left[1 - e^{-2 \pi \lambda(eA_-(x^+))}
\right] .
\end{eqnarray}
Integrating this last expression gives us the final result for the $++$ 
term,
\be
\langle \Omega \vert J^-_{5}(x^+,x^-,x^{\perp}) \vert \Omega \rangle_{++} 
\longrightarrow  -{e^2 E(x^+) B \over 8\pi^2} (x^- + L) \left[1 - e^{-2 \pi 
\lambda(eA_-)}\right]. \label{j5minusplus}
\ee
Note that this is not properly the infinite $L$ limit, but rather the two
leading terms --- one of which diverges linearly in $L$.

We pass now to the $--$ term. Reducing the transverse coordinates gives,
\begin{eqnarray}
\lefteqn{ \langle \Omega | J^-_{5}(x^+,y^+;x^-,x^{\perp}) | \Omega\rangle_{--}
= {eB \over 4\pi} {\partial \over \partial x^+} {\partial \over \partial y^+}
\int_0^{x^+} \!\!\!\!\! du \int_0^{y^+} \!\!\!\!\! dy {\cal{P}} \left({1 \over 
u-y}\right)} \nonumber \\*
& & \!\!\!\!\!\! \times \int_{-\infty}^{\infty} {dk^+ \over 2\pi} \frac{e^{-i
(k^+ + i/L) (x^- + L)}}{k^+ - eA_-(u) + i/L} \int_{-\infty}^{+\infty} {dq^+
\over 2\pi} \frac{e^{i (q^+ - i/L) (x^- + L)} e^{{i \over 2} m^2 \tau_{--}}}{
q^+ - eA_-(y) - i/L} . \quad
\end{eqnarray}
This can be recognized as $\frac{B}{2\pi}$ times expression (5.19) in ref. 
\cite{TW2}! So we can read off the result of the subsequent reductions from
expressions (5.25) and (5.28) of that paper,
\begin{equation}
\lim_{L \rightarrow \infty} \partial_- \langle \Omega | J^-_{5}(x^+,x^-,x^{
\perp}) | \Omega\rangle_{--} = {e^2 E(x^+) B \over 8\pi^2} \left[1 + e^{-2 \pi 
\lambda(eA_-)}\right] . 
\end{equation}
Integrating from $x^- = -L$ gives,
\begin{equation}
\langle \Omega | J^-_{5}(x^+,x^-,x^{\perp}) | \Omega\rangle_{--}
\longrightarrow  {e^2 E(x^+) B \over 8\pi^2} (x^- + L) \left[1 + e^{-2 \pi 
\lambda(eA_-)}\right] . \label{j5minusminus}
\end{equation}
Adding the $++$ terms (\ref{j5minusplus}) gives the final result for $J^-_5$,
\begin{equation}
\langle \Omega | J^-_5(x^+,x^-,x^{\perp}) | \Omega\rangle \longrightarrow
{e^2 E(x^+) B \over 4\pi^2} (x^- + L) e^{-2 \pi \lambda(p^+)}.
\end{equation}

As was the case for the vector current, the only divergence in the axial 
currents resides in $J^-_5$. Before computing the pseudo-scalar it is worth 
noting that in the massless limit the anomaly equation in 3+1 is simply,
\begin{equation}
\partial_{\mu} J^{\mu}_5 = {\alpha \over 4\pi} \epsilon^{\alpha \beta \mu\nu}
F_{\alpha \beta} F_{\mu\nu} = {e^2 E B \over 2 \pi^2} \label{zeromanomaly}
\end{equation}
Whereas our axial currents contain factors that are completely 
\textit{nonperturbative}, the limiting case satisfies (\ref{zeromanomaly}),
\be
\lim_{m \to 0} \left[ \partial_+ J^+_5 + \partial_- J^-_5 \right] = \lim_{m
\to 0} {e^2 E(x^+) B \over 2 \pi^2} e^{-2 \pi \lambda(p^+)} = {e^2 E(x^+) B
\over 2 \pi^2}. \label{masslesscheck}
\ee
Notice how (\ref{masslesscheck}) does not follow if the $--$ terms are 
suppressed. 

The only thing left to compute is the pseudo-scalar. We begin with the $++$ 
term,
\begin{eqnarray}
\lefteqn{ \langle \Omega \vert J_{5}(x^+,y^+;x^-,y^-) \vert \Omega \rangle_{++}
= - {e B m \over 8 \pi} e^{ie(x^- - y^-) \int_0^1 d\eta A_-\left(y^+ + \eta(x^+ 
- y^+)\right)}} \nonumber \\*
& & \times \left(\int_{-\infty}^0 \!\!\! - \!\! \int_0^{\infty} \right) {dp^+ 
\over 2\pi} \int_{-\infty}^{\infty}{dk^+ \over 2\pi} {e^{-i(k^+ + i/L) x^-}
\over k^+ - p^+ + \frac{i}{L}} \int_{-\infty}^{\infty} {dq^+ \over 2\pi} 
{e^{i(q^+ - i/L)y^-} \over q^+ -p^+ - \frac{i}{L}} \nonumber \\*
& & \times\left({1\over k^+ \! - \! eA_-(x^+) \! + \! \frac{i}{L}} - {1 \over 
q^+ \! - \! eA_-(y^+) \! - \! \frac{i}{L}} \right) e^{{i \over 2} m^2 
[\tau^*(0,y^+;q^+) - \tau_+]} , \\
& & \longrightarrow - {ieB \over 4\pi m} e^{ie(x^- - y^-) A_-} \left(
\int_{-\infty}^0 \!\!\! - \!\! \int_0^{\infty} \right) {dp^+ \over 2\pi} 
\nonumber \\
& & \times \left(\partial \over \partial x^+ \right) \int_{-\infty}^{\infty}
{dk^+ \over 2\pi} {e^{-i(k^+ +i/L) x^-} \over k^+ -p^+ + \frac{i}{L}} 
\int_{-\infty}^{\infty} {dq^+ \over 2\pi} {e^{i(q^+ -i/L) y^-} \over q^+ - p^+ 
- \frac{i}{L}} e^{{i \over 2} m^2 \tau_{++}} , \quad \\
& & \longrightarrow -{ieB \over 4\pi m} e^{-ieA_- \Delta_-} \partial_+ \left(
\int_{-\infty}^0 \!\!\!\! - \!\! \int_0^{\infty} \right) {dp^+ \over 2\pi} 
e^{ip^+ \Delta_-} e^{-2 \pi \lambda(p^+) \theta(p^+) \theta(eA_- - p^+)} , 
\qquad \\
& & = {ieB \over 8\pi^2 m} e^{-ieA_- \Delta_-} \partial_+ \Biggl[{i \over 
\Delta_-} \! + \! {ie^{ieA_- \Delta_-} \over \Delta_-} \! + \! \int_0^{eA_-} 
\!\!\!\!\!\! dp^+ e^{-2\pi \lambda(p^+) + ip^+ \Delta_-} \Biggl] , \quad \\*
& & \longrightarrow - {ie^2 A_-^{\prime}(x^+) B \over 8\pi^2 m}\left[1 -
e^{-2\pi \lambda(eA_-(x^+))}\right].
\end{eqnarray}
In these reductions we sequentially took $y^+ = x^+$, the large L limit, 
and then $y^- = x^-$. The final result is,
\begin{equation}
\lim_{ L \to \infty} \langle \Omega | J_5(x^+,x^-,x^{\perp}) | \Omega
\rangle_{++} = {i e^2 E(x^+) B \over 8 \pi^2 m} \left[1 - e^{-2 \pi 
\lambda(eA_-(x^+))} \right] . \label{j5plus}
\end{equation}

The $--$ term is perfectly regular at $x^+$ and $x^-$ coincidence, so we can
begin at coincidence,
\begin{eqnarray}
\lefteqn{ \langle \Omega \vert J_{5}(x^+,x^+;x^-,y^-) \vert \Omega \rangle_{--}
= -{ie B m \over 8\pi} \left({\partial \over \partial x^+} \right) \int_0^{x^+} 
\!\!\!\!\! du \int_0^{x^+} \!\!\!\!\! dy {i \over \pi} {\cal P}\left( {1 \over 
u-y}\right)} \nonumber \\
& & \times \int_{-\infty}^{\infty} {dk^+ \over 2\pi} \frac{e^{-i (k^+ + i/L)
x^-}}{k^+ - eA_-(u) + i/L} \int_{-\infty}^{\infty} {dq^+ \over 2\pi} \frac{
e^{i (q^+ - i/L) y^-} e^{{i \over 2} m^2 \tau_{--}}}{q^+ - eA_-(y) - i/L} , \\
& & \longrightarrow -{i eB \over 8 \pi^2 m} \left({\partial \over \partial x^+}
\right) \int_0^{eA_-(x^+)} dp^+ \left[1 - e^{-2 \pi \lambda(p^+)}\right] \\
& & = -{i e^2 A_-'(x^+) B \over 8\pi^2 m} \left[1 - e^{-2 \pi 
\lambda(eA_-(x^+))} \right] \; . \label{j5minus}
\end{eqnarray}
Combining (\ref{j5plus}) and (\ref{j5minus}) gives $J_{5}$,
\begin{equation}
\lim_{L \to \infty} \langle \Omega | J_5(x^+,x^-,x^{\perp}) | \Omega\rangle =
{i e^2 E(x^+) B \over 4 \pi^2 m} \left[ 1 - e^{-2 \pi \lambda(eA_-(x^+))} \; .
\right]
\end{equation}
With our results for the axial current, our divergence equation becomes,
\begin{eqnarray}
\lim_{L \to \infty} \langle \Omega | \partial_+ J^+_5 + \partial_- J^-_5 - 
2i m J_5 | \Omega \rangle = {e^2 E B \over 2 \pi^2}.
\end{eqnarray}
So the axial vector anomaly equation is satisfied and there are no 
non-perturbative corrections.

\section{Discussion}

This paper had three basic purposes. The first of these was to compute the 
positron creation probability and the vector current expectation values using 
operator solutions (\ref{plussolution},\ref{minussolution}) which are exact 
for any $L$. This is important because one cannot properly take the large $L$
limit --- or any other limit --- of an operator. The correct procedure is 
first to take the expectation value in the presence of some state and then 
take $L$ to infinity in the resulting $\comp$-number function. 

As in previous treatments \cite{TW1,TW2} pair creation in a homogeneous
electric field is a discrete and instantaneous event. For momentum $k^+$ it
occurs at the time $x^+ = X(k^+)$ such that $k^+ = eA_-(x^+)$. Electrons 
accelerate to the speed of light in the minus $z$ direction and leave the
light-cone manifold. In Section 4 we obtained the following probability for 
the appearance of a positron of momentum $k^+$, Landau level $n_-$ and spin 
$s$ is,
\begin{equation}
{\rm Prob}(k^+,n_-,s) = e^{-2 \pi \lambda(k^+,n_-,s)} \; ,
\end{equation}
where we define,
\begin{equation}
\lambda(k^+,n_-,s) \equiv {\frac12 m^2 + (2 n_- + 1 - 2 s) \vert \frac{eB}2
\vert \over \vert e E(X(k^+)) \vert } \; .
\end{equation}
It is reassuring that creation is more probable when the spin lines up with 
the magnetic field field ($s = + \frac12$).

In section 5 we obtained the following results for the nonzero currents,
\begin{eqnarray}
\Bigl\langle \Omega \Bigl\vert J^+(x^+,x^-,x^{\perp}) \Bigr\vert \Omega 
\Bigr\rangle & = & {e^2 B \over 4 \pi^2} \int_0^{e A_-} \!\!\!\!\! dk^+
e^{-\pi m^2 \over \vert e E(X(k^+)) \vert} \coth\left[{\scriptstyle {\pi 
B \over E(X(k^+))}} \right] \; , \\
\Bigl\langle \Omega \Bigl\vert J^-(x^+,x^-,x^{\perp}) \Bigr\vert \Omega 
\Bigr\rangle_{\rm ren} & = & {e^3 B E(x^+)\over 4 \pi^2} (x^- + L) 
e^{-\pi m^2 \over \vert e E(x^+) \vert} \coth\left[{\scriptstyle {\pi
B \over E(x^+)}} \right] . \qquad
\end{eqnarray}
We have removed the charge renormalization from $J^-$. Our results are
conserved, and they correctly reduce to the currents of ref. \cite{TW1} when 
$B = 0$. It may be that the extra magnetic field endows them with some
phenomenological significance. For whereas it is very difficult to maintain
large electric fields over long distances, there are many astrophysical 
sources which have large and quite extensive magnetic fields.

Our second object was to check the axial vector anomaly in $(3+1)$-dimensional
light-cone QED. Whereas an electric background suffices for checking the 
$(1+1)$-dimensional anomaly \cite{TW2}, increasing the dimensionality by 2 
requires the addition of a co-linear magnetic field. Although we chose this 
to be constant it seems feasible to consider more general backgrounds. For 
example, our solution (\ref{series_solution}) can be made valid for an $x^+$ 
dependent magnetic field $B(x^+)$ by the replacements,
\begin{eqnarray}
A_{\perp}(x^{\perp}) & \longrightarrow & A_{\perp}(x^+,x^{\perp}) = {B(x^+)
\over 2} (x^{2} \hat{x_1} - x^1 \hat{x_2}) , \\
{\cal U}(x^{\perp},\tau) & \longrightarrow & \exp\left[ -i \int_u^{x^+} du'
{{\cal H}[e A_{\perp}(u',x^{\perp})] \over k^+ - e A_-(u') + i/L} \right] \; .
\end{eqnarray}
This background entails transverse electric and magnetic fields,
\be
E^{\perp} = \frac1{\sqrt{8}} B'(x^+) (x^1 \widehat{x}^1 - x^2 \widehat{x}^2) 
\qquad , \qquad B^{\perp} = \frac1{\sqrt{8}} B'(x^+) (x^2 \widehat{x}^1 + x^1 
\widehat{x}^2) \;.
\ee
Although these make no contribution to the anomaly they do introduce an
interesting breaking of translation invariance in the transverse directions.

Our final purpose was to catalog the various disasters which ensue when the
operators at $x^- = -L$ are suppressed. One loses unitarity, current 
conservation and the axial vector anomaly. Not surprisingly, one also loses
renormalizability. For example, when we point-split on both $x^-$ and 
$x^{\perp}$ and then Hermitize, the $++$ part of the expectation value of 
$J^+$ is,
\begin{equation}
- {e \over 2\pi} \sum_{n_{\pm} , s} W^*_{n_{\pm}}(x^{\perp}) W_{n_{\pm}}(
x^{\perp} + \Delta^{\perp}) \int_0^{e A_-} \!\!\!\!\!\!\! dp^+ \Bigl[1 + 
e^{-2 \pi \lambda(p^+,n_-,s)} \Bigr] \cos[(p^+ \! - e A_-) \Delta^-] .
\end{equation}
The first term in the square brackets diverges quadratically like $\Vert 
\Delta^{\perp} \Vert^{-2}$. Yet the only counterterm QED allows for the 
current vector $J^{\mu}$ is $\partial_{\nu} F^{\nu\mu}$, which is only
nonzero for $\mu = -$ in our background.

What do these problems mean? There is a ``folk theorem'' to the effect that
anything one can see by studying the free theory with a nontrivial background 
must occur as well, in some way or another, for the interacting theory in a
trivial background. Of course the theory is fine if one includes the operators
on the $x^- = -L$ surface, but then much of the simplicity of light-cone 
quantum field theory is sacrificed. The best thing would be if the effects of
the extra operators could be subsumed into some simple extra interactions, at 
least for certain purposes. Quantifying the problem and deriving an appropriate
fix are the subject of on-going research.

Two extensions of this work seem worth making. The first is to compute the
one loop effective action with the addition of a static magnetic field. This
can no doubt be accomplished using the same techniques which worked for the
case of only an electric field \cite{FW}. It would be interesting to check
whether the Schwinger form persists in this larger class of backgrounds.

The second extension is to re-compute the large $L$ limits of the vector
currents under the assumption that $A_-(x^+)$ obeys the Maxwell equation,
\be
-A_-^{\prime\prime}(x^+) = \langle J^- \rangle
\ee
Since the term on the right hand side grows linearly with $L$, it is apparent
that the back-reacted vector potential must do the same. Our work of Section 5
assumed that $A_-(x^+)$ is fixed as $L$ goes to infinity.

\vskip 1cm
\centerline{\bf Acknowledgments}

It is a pleasure to acknowledge stimulating and informative conversations with
G. McCartor, T. N. Tomaras and N. C. Tsamis. This work was partially supported 
by DOE contract DE-FG02-97ER\-41029 and by the Institute for Fundamental 
Theory.

\end{document}